\documentclass[journal]{IEEEtran}

\usepackage{cite}
\usepackage{graphicx}
\usepackage{algorithm,algpseudocode}

\usepackage{amssymb}
\usepackage{mathrsfs}
\usepackage{amsmath}

\usepackage{stmaryrd}
\usepackage{pifont}
\usepackage{wasysym}

\usepackage{array}
\usepackage{url}
\usepackage{color}
\usepackage{caption}
\usepackage{multirow}
\usepackage{subcaption}
\usepackage{enumitem}

\newcommand{\plus}{\raisebox{.4\height}{\scalebox{.6}{+}}}
\newcommand{\minus}{\raisebox{.4\height}{\scalebox{.8}{-}}}

\algnewcommand{\LeftComment}[1]{\Statex \(\triangleright\) #1}

\makeatletter
\newcommand{\algmargin}{\the\ALG@thistlm}
\makeatother
\newlength{\whilewidth}
\algdef{SE}[parWHILE]{parWhile}{EndparWhile}[1]
{\parbox[t]{\dimexpr\linewidth-\algmargin}{%
		\hangindent\whilewidth\strut\algorithmicwhile\ #1\ \algorithmicdo\strut}}{\algorithmicend\ \algorithmicwhile}%
\algnewcommand{\parState}[1]{\State%
	\parbox[t]{\dimexpr\linewidth-\algmargin}{\strut #1\strut}}

\begin{document}

\title{Improving Viability of Electric Taxis by \\Taxi Service Strategy Optimization: \\ A Big Data Study of New York City}

\author{Chien-Ming Tseng, Sid Chi-Kin Chau and Xue Liu
\thanks{S. C.-K. Chau is with the Australian National University.  X. Liu is with McGill University. (Email: chi-kin.chau@cl.cam.ac.uk, xueliu@cs.mcgill.ca).}
\thanks{This paper appears in IEEE Transactions on Intelligent Transportation Systems (DOI:10.1109/TITS.2018.2839265).}
}

\maketitle

\begin{abstract}
Electrification of transportation is critical for a low-carbon society. In particular, public vehicles (e.g., taxis) provide a crucial opportunity for electrification. Despite the benefits of eco-friendliness and energy efficiency, adoption of electric taxis faces several obstacles, including constrained driving range, long recharging duration, limited charging stations and low gas price, all of which impede taxi drivers' decisions to switch to electric taxis. On the other hand, the popularity of ride-hailing mobile apps facilitates the computerization and optimization of taxi service strategies, which can provide computer-assisted decisions of navigation and roaming for taxi drivers to locate potential customers. This paper examines the viability of electric taxis with the assistance of taxi service strategy optimization, in comparison with conventional taxis with internal combustion engines. A big data study is provided using a large dataset of real-world taxi trips in New York City. Our methodology is to first model the computerized taxi service strategy by Markov Decision Process (MDP), and then obtain the optimized taxi service strategy based on NYC taxi trip dataset. The profitability of electric taxi drivers is studied empirically under various  battery capacity and charging conditions. Consequently, we shed light on the solutions that can improve viability of electric taxis. 
\end{abstract}


\section{Introduction}

Taxis are an important part of public transportation system, offering both flexibility of private vehicles and shareability of public transportation. In many cities around the world, there are usually a large of number of taxis, serving the ad hoc demands of commuters. Notably, taxis consume a large amount of fuel. For example, there are over 13,000 taxis operating in New York City, which totally travel over 1.46 billion kilometers each year\footnote{According to New York City taxi trip dataset in 2013 \cite{NYtaxi}.}, and consume over 86 million liters of gasoline. As a result, they emit over 242,900 metric tons of CO$_2$ per year\footnote{Estimated by assuming 67\% of New York Yellow taxis as hybrid vehicles and 33\% as ICE vehicles, as in 2016.}, which is equivalent to the amount of around 25,650 US households' average annual CO$_2$ emissions\footnote{The average annual CO$_2$ emission for US household is 9.5 metric tons~\cite{epa}.}.
A viable path toward a low-carbon sustainable society is to promote electrification of transportation, replacing internal combustion engine (ICE) vehicles by more environment-friendly and energy-efficient electric vehicles (EVs). Electrification of private vehicles faces many obstacles, such as cost-effectiveness, availability of home charging infrastructure and users' perception. However, electrification of public vehicles (e.g., buses, taxis) would be subject to fewer concerns, with even a greater potential impact than that of private vehicles. First, public vehicles are used more frequently, whose electrification can effectively reduce greenhouse gas emissions. Second, public vehicles are likely to park in common facilities, facilitating the installation of charging stations. Third, public vehicles generally have shorter life cycles due to frequent usage, and hence, are more ready to be replaced.

Major cities worldwide are introducing plans to phase out conventional ICE public vehicles for electric vehicles. For example, Chinese government has initiated several programs to promote electrification of public vehicles for air pollution mitigation \cite{BYDtaxi}.  Electric taxi programs were launched in Shenzhen (in 2010) and Beijing (in 2014) to convert taxis to electric vehicles, along with the installation of sufficient EV parking lots and fast charging points. In these programs, the government also offer subsidies to taxi operators. Singapore government plans to roll out a total of 1,000 electric cars to be supported by 2,000 charging points across the city by 2020.

Nonetheless, unlike buses, taxis are often operated as private businesses. Adoption of electric taxis critically depends on the willingness of taxi drivers to switch to electric taxis from conventional ICE taxis. However, it is not clear whether taxi drivers are willing to do so. Despite the initiatives from the governments, there are notable shortcomings of electric taxis:
\begin{enumerate}

\item {\bf Constrained Driving Range}: One of the barriers preventing wide adoptions of EVs is a shorter driving range. With increasing battery capacity, the driving range has been extended to more than 200 kilometers in production EVs such as Chevrolet Bolt. Generally, the driving ranges of production EVs are sufficient for daily commutes of personal purposes. However, a longer driving range is normally required by logistic vehicles and taxis (e.g., more than 300 kilometers). The driving range of high-end Tesla (as in 2017) may suffice to meet the required driving distance, but are too costly for practical taxis.

\item {\bf Long Recharging Duration}: Recharging the battery of EVs can take considerable time. For example, charging Nissan Leaf with 30 kwh battery capacity can take up to 4 hours using mode 3 charging, or half an hour using fast DC charging (without considering queuing delay). Taxis traveling long distances are likely to take more than an hour for recharging between shifts, which is significantly longer than ICE taxis with faster refilling of gasoline.

\item {\bf Limited Charging Stations}: Todays, the number of charging stations are few. Also, some of charging stations are reserved for specific models or brands with proprietary connectors. The expansion of charging stations is hampered by electrical infrastructure in certain regions. As a result, electric taxi drivers always need sufficient reserve battery capacity in order to be able to return to certain known charging stations, in case of emergence.

\item {\bf Low Gas Price}: Nowadays, the oil price has come down considerably from historic heights. This reduces the incentive to adopt EVs, as the gasoline is relatively affordable, despite cheaper and cleaner electricity sources. Unless carbon tax is introduced to mitigate greenhouse gas emissions, gasoline ICE vehicles are still perceived as cost-effective by the public in general.

\end{enumerate}

These shortcomings are likely to dissuade taxi drivers from adopting electric taxis. Particularly, it is not easy to operate a taxi under the constraints of shorter driving range and limited charging stations, in comparison with conventional taxis. In fact, it has been reported in media that taxi drivers tended to shun electric taxis. Without taxi drivers' participation, it is futile to promote electric taxis. Therefore, it is important to provide a viability analysis of electric taxis. Such an analysis can also be used as a basis to determine proper governmental subsidies for electric taxis to promote their adoptions.

In this paper, we identify that a key problem of adopting electric taxis is the ineffective service strategies practiced by today's taxi drivers. In fact, we show that properly optimized taxi service strategies will not suffer from the shortcomings of electric vehicles. Therefore, there is a need to provide an intelligent recommender system to assist taxi drivers to improve their taxi service strategies, and hence, to increase their willingness to switch to electric taxis. In particular, there is a popular trend of ride-hailing mobile apps, which facilitates the computerization and optimization of taxi service strategies, and provide an opportunity of integrating computer-assisted optimized decisions of roaming and navigation to taxi drivers.

\subsection{Modeling Taxi Service Strategy by MDP}

The net revenue of a taxi driver (i.e., the revenue from taxi fares minus energy costs) is determined by his/her service strategy of passenger searching and efficiency of passenger delivery. For example, skilful taxi drivers can identify the popular spots for potential passengers, and deliver passengers efficiently by choosing faster routes. Note that the service strategies of taxi drivers can be effectively optimized by utilizing a large historical taxi trip dataset for demand prediction.

To optimize taxi service strategies for electric (or ICE) taxis, we first model computerized taxi service strategy by {\em Markov Decision Process} (MDP). MDP is a general framework for optimizing sequential decision process in the presence of uncertainty. In summary, we denote a Markov state as the time and location (and possibly battery state) of a taxi, and an action as the driver's decision to travel to the next location (and possibly recharging operations). At each location, there is a probabilistic transition to another location. The transition is determined by a random event of passenger pick-up. The uncertainty in taxi service strategy is the pick-up location and destination of a passenger, which can be estimated by a historical taxi trip dataset.

This MDP model facilitates the optimization of computerized taxi service strategies by providing computer-assisted decisions to taxi drivers. Since human taxi service strategies are inherently inefficient,  optimizing computerized taxi service strategies can potentially improve the net revenues of taxi drivers, particularly in presence of constraints of driving range and charging stations. Computerized taxi service strategies are becoming more feasible, because the increasing adoption of ride-hailing mobile apps, which  facilitates the integration of computerized taxi service strategies in a recommender system for taxi drivers using real-time data analytics from historical taxi trip dataset. In this paper, we obtain the optimal policy of MDP that maximizes the revenue of a taxi driver based on New York City taxi trip dataset, and study the profitability of electric taxi drivers  under various conditions of battery capacity and charging modes.

\subsection{Summary}

Our contributions in this paper are summarized as follows:
\begin{enumerate}

\item We formulate an MDP to model computerized electric taxi service strategies, with explicit consideration of constraints of EVs, such as battery capacity and locations of charging stations.

\item We obtain the optimal policy of the MDP based on a big data study using a large dataset of real-world taxi trips in New York City.

\item We study the impact of factors such as battery capacity and charging modes, and locations of charging stations on the net revenues of electric taxi drivers.

\item We project our study to understand the benefits of a wider adoption of electric taxis (up to 1000 taxis).

\end{enumerate}

\section{Background}

\subsection{Related Work}

Analyzing taxi trip dataset has been considered by several research papers in the subjects of knowledge discovery and cloud-based intelligent transportation systems \cite{cloudthink15}. One of the popular topics is the profit/revenue improvement for taxi drivers by developing a recommender system for assisting the drivers to find passengers more efficiently. The basic idea is to identify the good taxi service strategies. Several characteristics of taxi service strategies are reported in \cite{daqing2015taxistrategy}. Their study shows that searching passengers near the drop-off location of previous passengers results in a higher revenue. They also found that better taxi drivers can deliver the passengers efficiently by choosing a uncongested route.
Furthermore, GPS mobility trace from taxis can be used to predict future traffic conditions and optimize the route selections \cite{siyuan2015adacongestion}. Also, community detection has been applied to the mobility trace to reveal potential similar passengers' travel patterns, as for social recommendation \cite{siyuan2017trajmdl} and improving transportation services \cite{juanjuan2017smartcard}.

Other studies focus on the specific methods for improving the profit/revenue of the taxi drivers. One approach in \cite{yuan2011findpass} shows that experienced taxi drivers usually waits for passengers at specific locations, and they are usually aware of particular events like train arrivals or ending times of movies.

\begin{figure*}[!htb]
    \begin{subfigure}[h]{0.36\textwidth}
        \includegraphics[width=\textwidth]{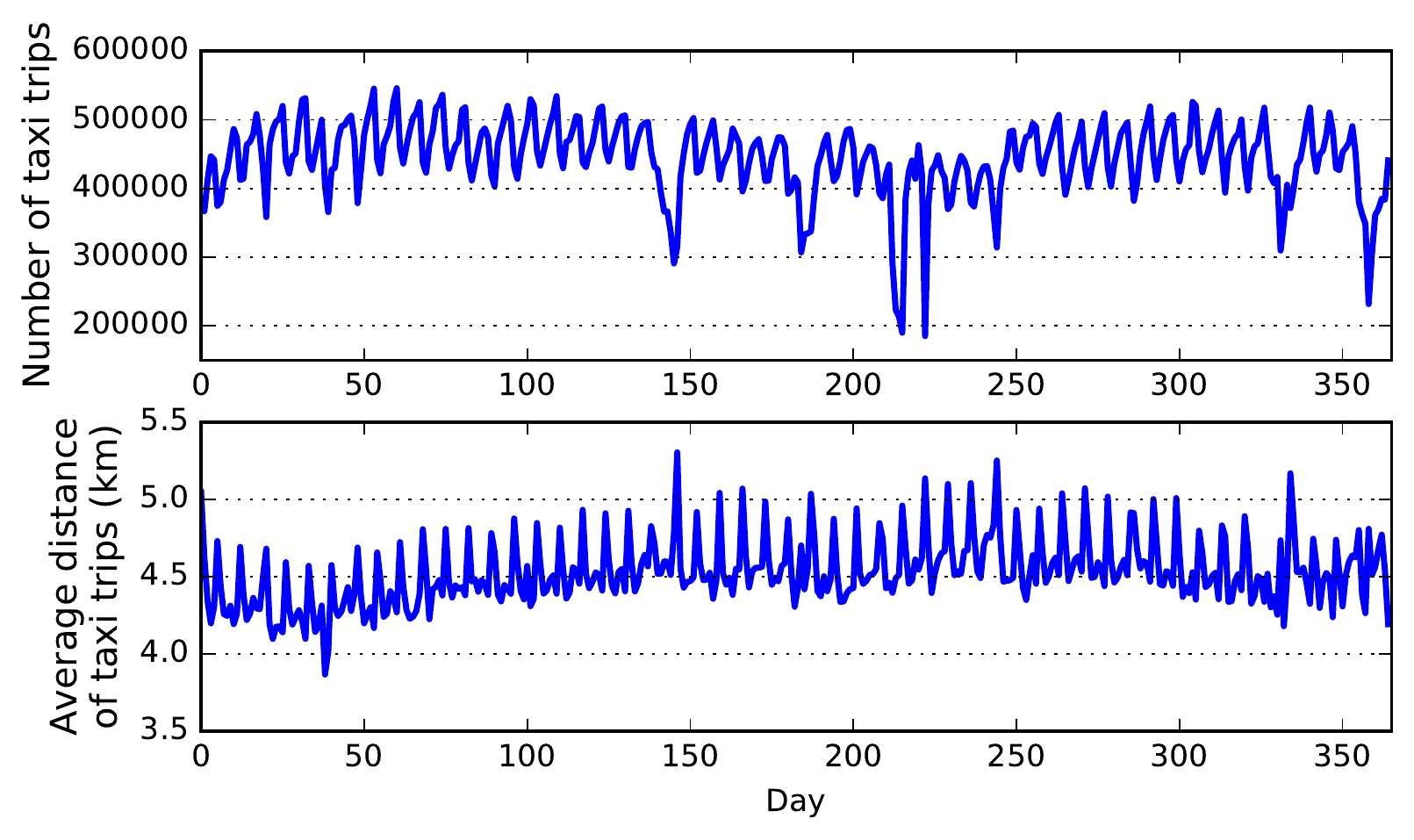} 
        \caption{Num. of trips and average trip distance of NYC taxi trip dataset.}
        \label{fig:NYdata}
    \end{subfigure}
    ~
    \begin{subfigure}[h]{0.275\textwidth}
        \includegraphics[width=0.96\textwidth]{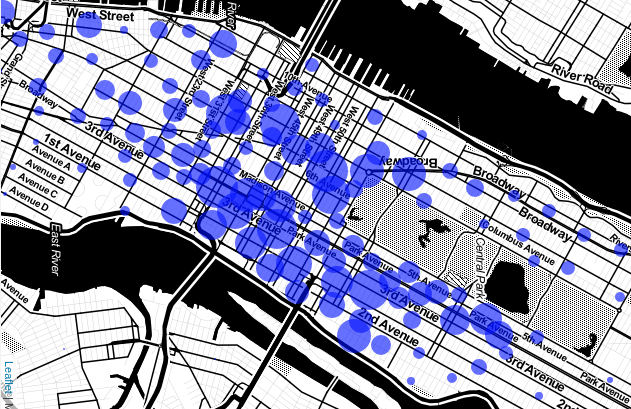} 
        \caption{Pick-up locations in NYC based on k-means clustering.}
        \label{fig:NYpic}
    \end{subfigure}
    ~
    \begin{subfigure}[h]{0.325\textwidth}
        \includegraphics[width=1\textwidth]{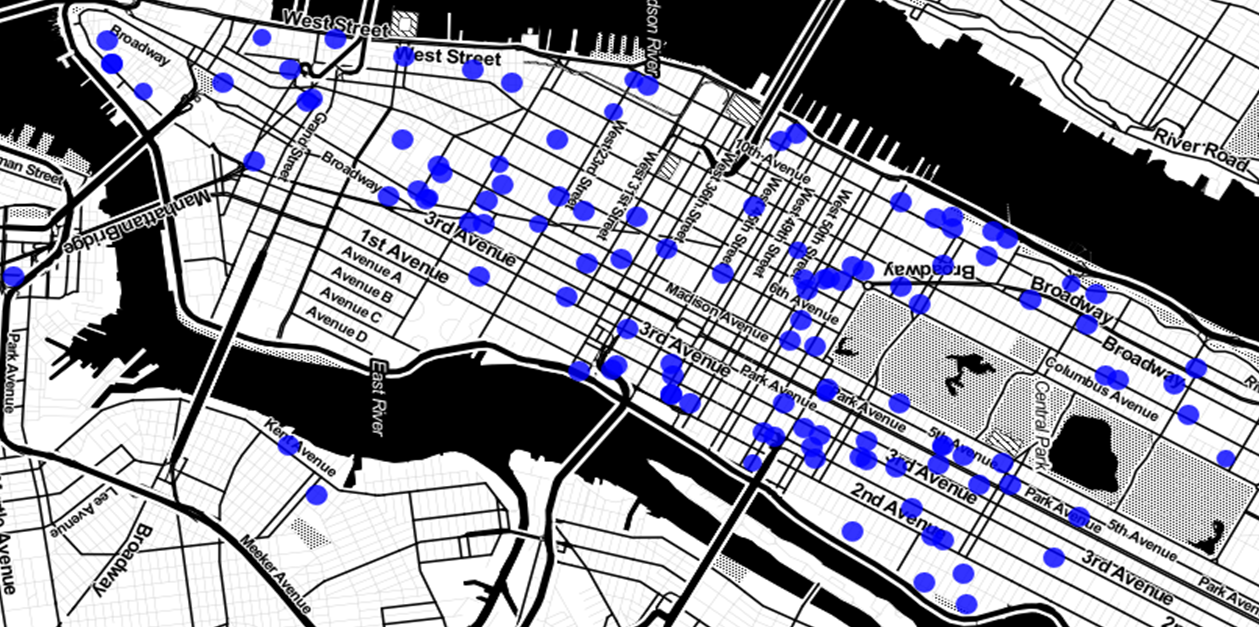}
        \caption{Charging stations in NYC.}
        \label{fig:NYchg}
    \end{subfigure}
    \caption{Overview of NYC taxi trip dataset and locations of charging stations.}
\end{figure*}

Instead of recommending separate pick-up locations, a better approach is to maximize the revenue by selecting a route of a sequence of likely pick-up locations at different times. The top-k profitable driving routes can be computed based on a route network with revenues and pick-up probabilities from historical taxi trip data in \cite{qu2014effrecommend}. To select an optimal route with appropriate actions, Markov Decision Process (MDP) is used to maximize the associated revenue in \cite{rong2016taximarkov}. The optimal policy of MDP is determined to improve the taxi driver's service strategy. The method of MDP is significantly extended in this paper to consider the constraints of EVs, such as battery capacity and locations of charging stations. Our preliminary study \cite{evsys17} uses a simplified model, whereas this paper presents a more realistic model and a more extensive analysis.

For EVs, limited driving range is a barrier preventing wide adoption. Therefore, the estimation of driving range for EVs has been studied in a number of research papers. The driving range of EVs is highly affected by driving speed and motor efficiency.  A black-box model is widely used in the literature to predict the energy consumption of  EVs and plug-in hybrid EVs (PHEVs) \cite{cmtseng2017dte,ckt2017phevopt}.  Such a black-box model is used in this paper to estimate the energy consumption of electric taxis.

There are other studies that investigated the viability of deploying electric taxis. For example, the return on investment (ROI) for taxi companies transitioning to EVs was studied in \cite{carpenter2014evroi}, which considers the mobility trace of yellow cabs in San Francisco. The prior studies usually assumed that electric taxi drivers will adopt the same service strategies as driving a conventional ICE taxi. On the contrary, our study allows distinctive optimized service strategies for electric taxi drivers, taking into account that EVs have different operating constraints than conventional ICE vehicles.

\subsection{New York City Taxi Trip Dataset}

We describe the taxi trip dataset of New York City (NYC) of 2013 that is used in our study. In the following, we list the attributes of dataset that are used in our study. For each data record (i.e., a trip), it is composed of following attributes:
\begin{itemize}

\item Taxi ID (also known as medallion ID)

\item Trip distance and duration

\item Times of pick-ups and drop-offs of passengers

\item GPS locations of pick-ups and drop-offs of passengers

\end{itemize}
We summarize the information of taxi trip dataset in Table \ref{tab:NYoverviewTab}.

\begin{table}[!htb]
\centering
\caption{New York City taxi trip dataset in 2013.} \label{tab:NYoverviewTab}
\begin{tabular}{@{}c|c@{}}
\hline
\hline
Attribute & Quantity  \\
\hline
Num. of medallions (i.e., rights to operate a taxi) & 13437 \\
Annual average traveled distance per taxi &  112,600 km \\
Total num. of trips & 175M \\
Average num. of trips per day & 450,000 \\
Average trip distance & 4.2 km \\
\hline \hline
  \end{tabular}
\end{table}

The numbers of taxi trips of NYC dataset on different days of 2013 are depicted in Fig.~\ref{fig:NYdata}. There are about 450K trips per day and the average trip distance is around 4.2 km. Fig.~\ref{fig:NYpic} displays the pick-up locations on January 16  at 8-9 AM. The k-means algorithm is employed to cluster the pick-up locations by 200 clusters. The sizes of circles indicate the number of pick-up locations. We observe most of pick-ups occur in Midtown Manhattan. Finally, Fig.~\ref{fig:NYchg} displays the locations of charging stations in NYC \cite{nychg} that potentially recharge electric taxis.

\section{Markov Decision Process Model} \label{sec:mdp}

In this work, we extend  the Markov Decision Process (MDP) framework in \cite{rong2016taximarkov} to model the computerized service strategy of an electric taxi. MDP facilitates the formulation of computerized taxi service strategies, which can be implemented in a recommender system for taxi drivers.
In general, a MDP comprises of a set of states  and a set of possible actions at each state. Each action transfers the current state to a new state with a probability and a reward. The objective is to find the optimal actions in the corresponding states that maximize the expected total reward.

\subsection{States and Actions}

First, we explain the states and actions of the MDP in our setting. A state for an electric taxi is described by three parameters: current time, current location and battery state, as explained as follows.
\begin{itemize}

\item {\em Current Time}:
We consider discrete timeslots. One minute is used as the interval of a timeslot.

\item {\em Current Location}: We consider the locations represented by the nearest junctions, instead of the absolute locations.
A road network is constructed using OpenStreetMap (OSM) junction data. Each pick-up or drop-off location is assigned to the nearest junction in OSM.  Let ${\cal N}$ be the set of all junctions.

\item {\em Battery State}:
We consider discrete levels of state-of-charge of battery of the electric taxi. The feasible battery state should be within the range $[\underline{B}, \overline{B}]$.

\end{itemize}
We denote the location of a taxi at time $t$ by $S(t)$, and the battery state by $B(t)$.

The allowable actions at the current junction are the neighbors of the junction in the road network, and the recharging duration, if the electric taxi is subject to recharging at this junction. We denote an action from junction $i$ to junction $j$ with recharging duration $\tau$ at $i$ by $A = (i \to j, \tau)$, where $i$ and $j$ are  neighbors in the road network.

\subsection{State Transition and Objective Function}

\begin{figure}[htb!] 
        \includegraphics[width=0.4\textwidth]{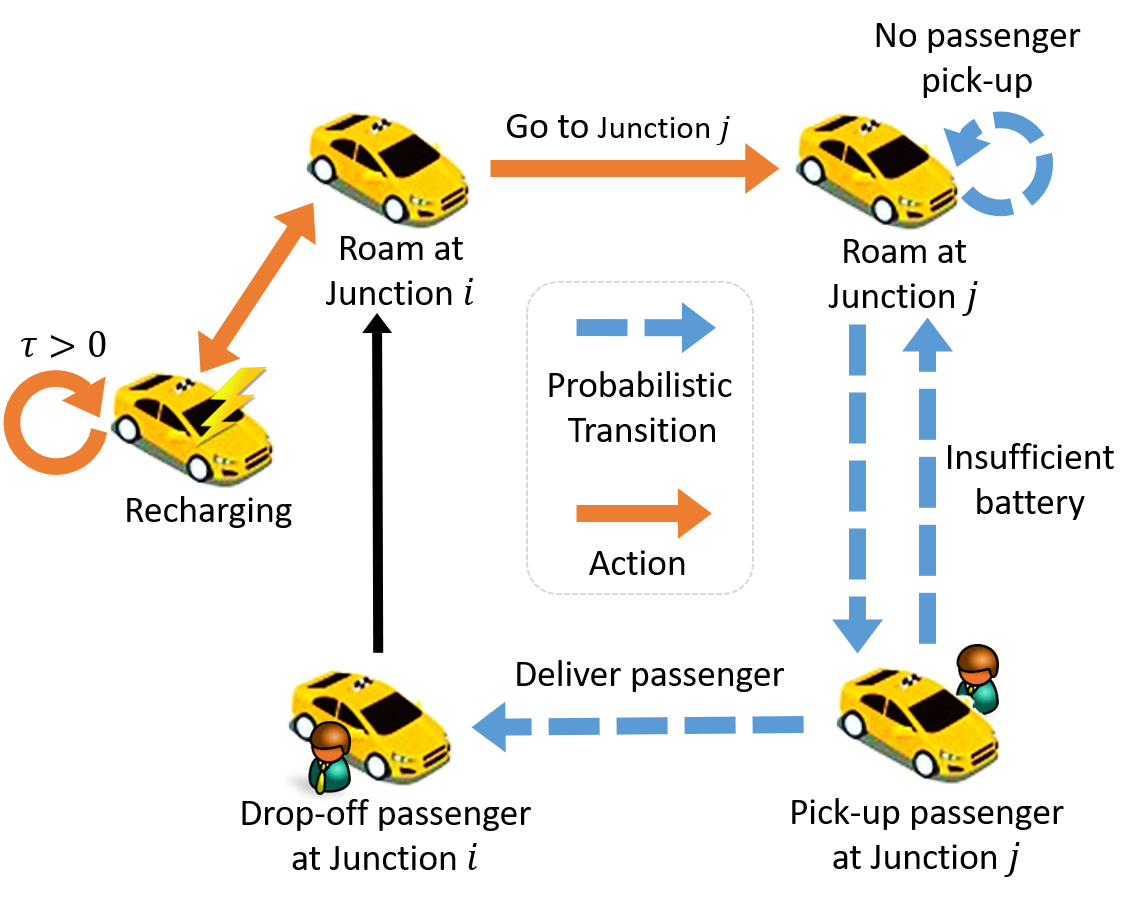} 
        \caption{An illustration of MDP for the computerized service strategy of an electric taxi.}
        \label{fig:MarkovChain}
    \end{figure}

The basic idea of the MDP for computerized taxi service strategy is illustrated in Fig.~\ref{fig:MarkovChain}.  Assuming the current location is $i$, action $A  =  (i \to j, \tau)$ is taken. The next location will be $j$ after recharging for a duration $\tau$ at $i$. When entering junction $j$,  there is a probability of not picking up any passenger, after which the taxi driver will make another action. On the other hand, there is a probability of picking up a passenger, with a random destination. The taxi driver will decide if the current battery state is sufficient to deliver the passenger to the respective destination, or the trip is discarded. The detailed descriptions of MDP are provided in the following.

First, we define several parameters for the MDP as follows.
\begin{itemize}

\item $P^{\tt{p}}_{t}(i)$: The probability of successfully picking up a passenger at junction $i$ at time $t$.
\item $P^{\tt{d}}_{t}(i,j)$: The probability of a passenger commuting from junction $i$ to junction $j$ at time $t$.
\item $T^{\tt{a}}_{t}(A)$: The required time (mins) for executing action $A$.
\item $T^{\tt{t}}_{t}(i,j)$: The required traveling time (mins) from junction $i$ to junction $j$ at time $t$.
\item $E^{\tt{e}}_{t}(i,j)$: The required energy consumption (kW) from junction $i$ to junction $j$ at time $t$.
\item $F_{t}(i,j)$: The net revenue of transporting passengers from junction $i$ to junction $j$, which is calculated based on the fare rule of New York taxi and the respective energy costs. There are various surcharges in different times and days, and hence, the net revenue is time-dependent.
\item $U^{\tt{a}}_{t}(i,j)$: The energy cost from junction $i$ to $j$ at time $t$.

\end{itemize}
Note that some of these parameters (e.g., $P^{\tt{p}}_{t}(i)$, $P^{\tt{d}}_{t}(i,j)$, $T^{\tt{t}}_{t}(i,j)$, $E^{\tt{e}}_{t}(i,j)$) can be estimated from the taxi trip dataset, which will be discussed in the subsequent section.

Next, we formulate a recurrent equation for describing the MDP, namely, Eqn.~(\ref{eq:obj}) (as illustrated in Fig.~\ref{fig:eq1}).

If the current location is $S(t) = i$, after action $A  =  (i \to j, \tau)$ has been taken, the next location will be $S(t') = j$, where $t' \triangleq  t+T^{\tt{a}}_{t}(A)$. The required time of the action $T^{\tt{a}}_{t}(A)$ is computed as follows:
\begin{enumerate}

\item If recharging duration $\tau=0$, the taxi directly goes to junction $j$. The required time of action is given by $$T^{\tt{a}}_{t}(A) = T^{\tt{t}}_{t}(i,j)$$

\item If recharging duration $\tau>0$, before driving to junction $j$, the taxi first goes to the nearest charging station $r(i)$ to recharge the electric taxi. The required traveling time is $T^{\tt{t}}_{t}(i,r(i))$ to travel to charging station $r(i)$. Then the electric taxi is recharged for $\tau$ duration and next goes from charging station $r(i)$ to junction $j$, whose required traveling time is $T^{\tt{t}}_{t\plus T^{\tt{t}}_{t}(i,r(i))\plus\tau}(r(i),j)$.
Thus, the total required time of action is given by $$T^{\tt{a}}_{t}(A) = T^{\tt{t}}_{t}(i,r(i))+\tau+T^{\tt{t}}_{t\plus T^{\tt{t}}_{t}(i,r(i))\plus\tau}(r(i),j)$$

\end{enumerate}

Note that if the state-of-charge of battery is insufficient, certain actions are infeasible (e.g., driving to a distant location to pick up passengers). Therefore, an action needs to consider the required energy consumption that can be supported by the current battery state. If the current battery state is $B(t) = b$, after action $A  =  (i \to j, \tau)$ has been taken, the new battery state at $j$ will be $B(t') = b' \triangleq \min\{ b + \tau C-E^{\tt{e}}_{t}(i,j)-E^{\tt{e}}_{t}(i,r(i)), \overline{B}\}$, where $C$ is the charging rate, and $t' \triangleq  t+T^{\tt{a}}_{t}(A)$.

\begin{figure*}[!t]
\begin{equation}
\label{eq:obj}
\begin{aligned}
 R^*[t,i,b,A] = & \Big(1 -  P^{\tt{p}}_{t'} (j) + \sum_{k \in {\cal N}:E^{\tt{e}}_{t}(j,k)\plus E^{\tt{e}}_{t}(k,r(k))\plus\underline{B}> b'} P^{\tt{p}}_{t'}(j)P^{\tt{d}}_{t'}(j,k) \Big) \cdot R^*[t',j,b'] \\
& + \sum_{k \in {\cal N}:E^{\tt{e}}_{t}(j,k)\plus E^{\tt{e}}_{t}(k,r(k))\plus\underline{B}\le b'} P^{\tt{p}}_{t'}(j) P^{\tt{d}}_{t'}(j,k) \cdot\Big(F_{t'}(j,k) + R^*\big[t'\plus T^{\tt{t}}_{j,k,t'},k, b'\big] - E^{\tt{e}}_{t'}(j,k)\Big)-U^{\tt{a}}_{t}(i,j)
\end{aligned}
\end{equation}
\end{figure*}

At junction $j$, there are three possible state transitions:
\begin{enumerate}

\item[(C1)] The taxi successfully picks up a passenger at junction $j$ (say, with destination $k$) and $B(t')$ is sufficient to deliver the passenger to junction $k$ and then to the nearest charging station $r(k)$, if necessary. For each $k$, the probability is  $P^{\tt{p}}_{t'}(j) P^{\tt{d}}_{t'}(j,k)$, subject to the constraint $E^{\tt{e}}_{t}(j,k)+E^{\tt{e}}_{t}(k,r(k))+\underline{B}\le b'$, such that the resultant battery state is always larger than the minimal $\underline{B}$. Hence, denote the probability of picking up a passenger by probability $P^{\tt p}_{t'}(j)P^{\tt s}_{t'}(j)$, where $P^{\tt s}_{t'}(j)$ is the probability that the destination of passenger is reachable for the taxi under battery constraint, and is computed by
$$
P^{\tt s}_{t'}(j)=\sum_{k \in {\cal N}:E^{\tt{e}}_{t}(j,k) + E^{\tt{e}}_{t}(k,r(k)) + \underline{B} \le b'} P^{\tt{d}}_{t'}(j,k)
$$

\item[(C2)] The taxi successfully picks up a passenger at junction $j$, but $B(t')$ is insufficient to deliver the passenger to junction $k$ and then to the nearest charging station $r(k)$. The total probability of such a case is $$\sum_{k \in {\cal N}:E^{\tt{e}}_{t}(j,k) \plus  E^{\tt{e}}_{t}(k,r(k)) \plus \underline{B}> b'} P^{\tt{p}}_{t'}(j)P^{\tt{d}}_{t'}(j,k)$$

\item [(C3)] The taxi cannot successfully pick up a passenger at junction $j$. The  probability is $1-P^{\tt{p}}_{t'}(j)$.

\end{enumerate}

Note that the probability that the taxi does not deliver any passenger (including (C2) and (C3)) is $1-P^{\tt p}_{t'}(j)\cdot P^{\tt s}_{t'}(j)$. The complement of $P^{\tt s}_{t'}(j)$, i.e., $1-P^{\tt s}_{t'}(j)$ is given by
$$
1-P^{\tt s}_{t'}(j)\triangleq\sum_{k \in {\cal N}:E^{\tt{e}}_{t}(j,k) + E^{\tt{e}}_{t}(k,r(k)) + \underline{B} > b'} P^{\tt{d}}_{t'}(j,k)
$$
Hence, we obtain
\begin{align*}
& 1-P^{\tt p}_{t'}(j)\cdot P^{\tt s}_{t'}(j) \\
 = & 1-P^{\tt p}_{t'}(j)+P^{\tt p}_{t'}(j)\cdot (1-P^{\tt s}_{t'}(j)) \\
= & 1 -  P^{\tt{p}}_{t'} (j) + \sum_{k \in {\cal N}:E^{\tt{e}}_{t}(j,k)\plus E^{\tt{e}}_{t}(k,r(k))\plus\underline{B}> b'} P^{\tt{p}}_{t'}(j)P^{\tt{d}}_{t'}(j,k)
\end{align*}

For (C1),  the taxi driver will receive a fare of amount $F_{t'}(j,k)$, and the next location of the taxi becomes $S(t'+T^{\tt{t}}_{j,k,t'})=k$. For (C2) and (C3),  the taxi driver will not receive any fare,  and will decide to drive to another location or possibly recharge the taxi.

The objective of the MDP is to maximize the total expected net revenue. Note that the net revenue of the action is the received fare minus the energy cost of the action. The expected net revenue for an action $A = (i \to j, \tau)$ at state $(t,i,b)$ is denoted by $R^*[t,i,b,A]$, which can be computed recurrently in Eqn.~(\ref{eq:obj}), where
\begin{itemize}
\item $t' =  t+T^{\tt{a}}_{t}(A)$ and $b' = \min\{ b + \tau C, \overline{B}\}$.
\item $R^*[t,j,b'] = \max_{A} R^*[t,j,b',A]$ is the maximal expected net revenue in state $(t,j,b')$ over all possible actions.
\item $U^{\tt{a}}_{t}(i,j)$ is the energy cost, as computed as follows:
\begin{enumerate}

\item If recharging duration $\tau=0$, the taxi directly goes to junction $j$. The energy cost is
$U^{\tt{a}}_{t}(i,j)=E^{\tt{e}}_{t}(i,j) \cdot U$, where $U$ is the unit price, such that $U=$20 cent/kWh for electricity and $U=$2.5 USD\$/gallon for gasoline.

\item If recharging duration $\tau>0$, the taxi goes to the nearest charging station $r(i)$ to recharge the electric taxi at charging rate $C$.  The energy cost of the action is given by
$$U^{\tt{a}}_{t}(i,j)=\big(
E^{\tt{e}}_{t}(i,r(i))+E^{\tt{e}}_{t\plus T^{\tt{t}}_{t}(i,r(i))\plus \tau}(r(i),k)+\tau \cdot C\big) \cdot U$$
\end{enumerate}

\end{itemize}

 \begin{figure}[htb!] \center
        \includegraphics[width=0.51\textwidth]{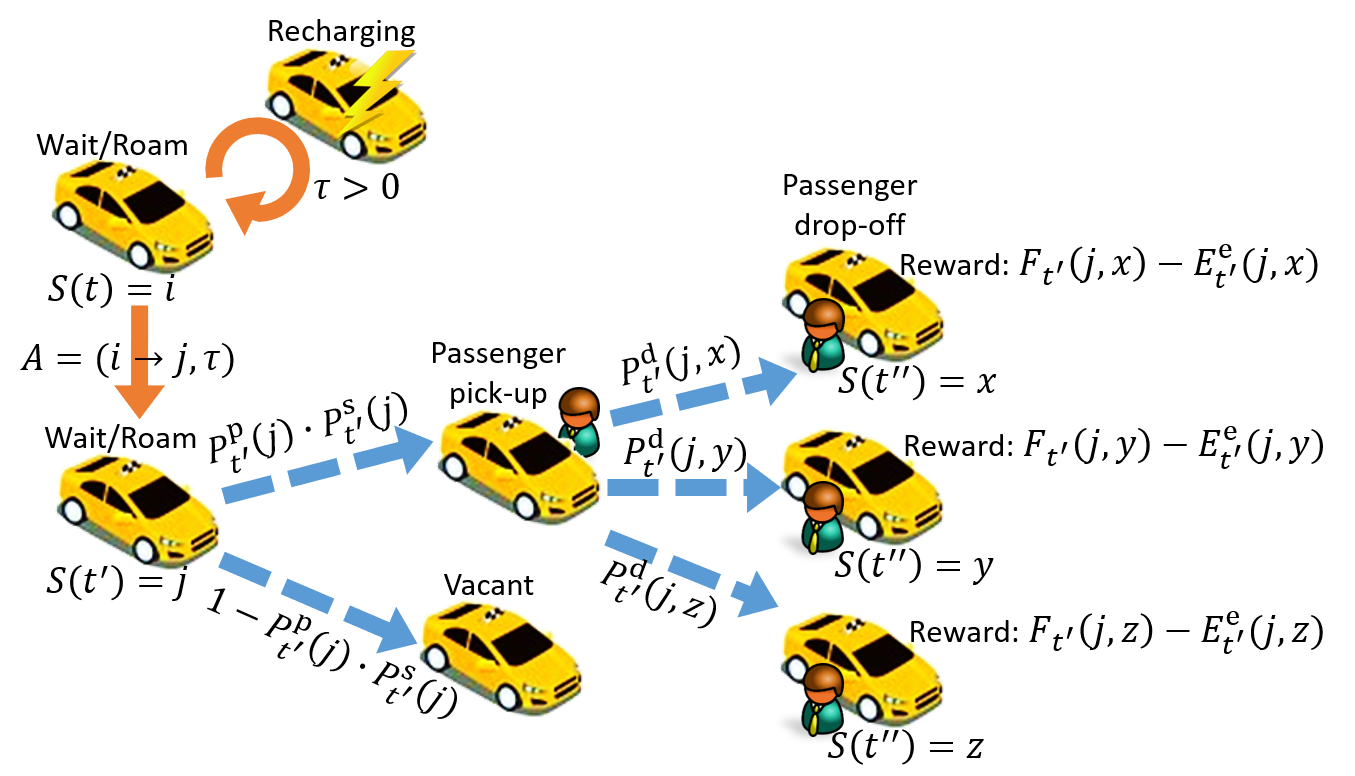} 
        \caption{An illustration for the  recurrent equation Eqn.~(\ref{eq:obj}).}
        \label{fig:eq1}
    \end{figure}

We seek to devise an optimal policy $\pi$ for the MDP that maximizes the expected net revenue:
\begin{align}
 \pi(t,S(t),B(t))= \ &  \arg\max_{A} R^* \big[t, S(t), B(t), A \big]
\end{align}

To obtain the optimal policy for the MDP, one can use dynamic programming.
The dynamic programming algorithm starts from the last timeslot and then works backwards to the beginning timeslot. For example, to solve the optimal policy for a morning shift, the algorithm starts to solve the maximal expected net revenue at the end of shift, and works backwards.

\section{Markov Decision Process Parameters} \label{sec:param}

In this section, we estimate several parameters of MDP (e.g., $P^{\tt{p}}_{t}(i)$, $P^{\tt{d}}_{t}(i,j)$, $T^{\tt{t}}_{t}(i,j)$, $E^{\tt{e}}_{t}(i,j)$) from NYC taxi trip dataset.

\subsection{Driving Speed Network}

First, we construct a driving speed network from the NYC taxi trip dataset, for the following purposes:
\begin{enumerate}

\item To estimate the traveling time from each junction to the nearest charging station.

\item To estimate the energy consumption of a taxi for a trip.

\end{enumerate}
Note that traveling time and driving speed are time-dependent parameters, since they are highly affected by traffic condition, which is estimated from historical trip data. For example, the traveling time between the same pair of junction $i$ and junction $j$ will be higher in office hours and much lower at midnight.

\begin{figure}[!htb]
    \center
	\includegraphics[width=0.47\textwidth]{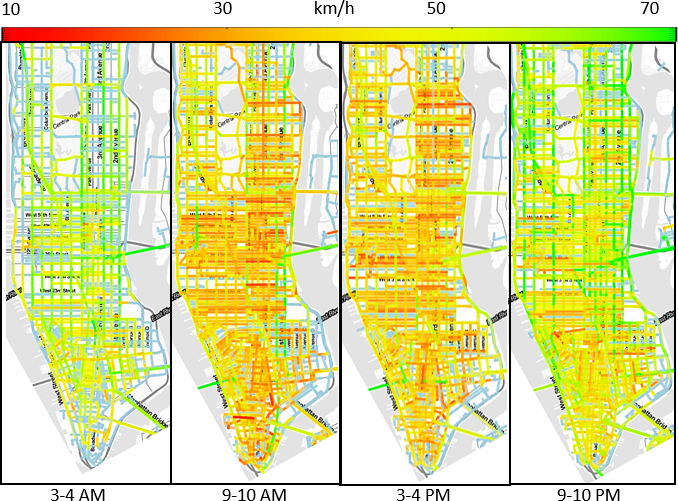} 
    \caption{Visualizations of driving speed networks.}
    \label{fig:trafficnetwork}
\end{figure}

The first step of constructing the driving speed network is to determine the driving path of a taxi. Spatialite \cite{spatialite} is used to calculate the shortest path for each pair of pick-up and drop-off locations. Spatialite utilizes OpenStreetMap (OSM) data. A resulting path comprises a list of edges (i.e., segments) described by two junctions. We then compare the recorded trip distance in the taxi trip dataset to the computed shortest path distance. If the difference is greater than 300 meters, the record is discarded since the driver is likely to take other route. For each computed path, the segments of a path are labeled with the average speed using recorded traveling time and distance.
We can obtain the average speed for each taxi trip record. Each segment has several average speeds by different trips. We select the highest speed to represent the driving speed of the segment, since this is usually the speed with minimal obstacles.
Driving speed networks at different times are visualized in Fig.~\ref{fig:trafficnetwork}. We observe there is relatively more congested traffic in 9 to 10 AM or 4 to 5 PM. 

Given the driving speed network, we can estimate the driving time from the network. We can also estimate the idling time of each trip by subtracting the estimated driving time from the recorded traveling time. The detailed steps for calculating the idling time are described as follows:
\begin{enumerate}

\item Average traveling time $T^{\tt{t}}_{t}(i,j)$: There may be several trips start from junction $i$ to junction $j$. However, their traveling times may be slightly different. We average the traveling time of these trips.

\item Driving time $T^{\tt{d}}_{t}(i,j)$: The shortest path from junction $i$ to junction $j$ is determined by Spatialite. Then, the driving time in each segment is computed by its distance and the driving speed from the driving speed network.

\item Idling time $T^{\tt{i}}_{t}(i,j)$: The idling time of a trip is obtained by subtracting the driving time from the average traveling time, $T^{\tt{i}}_{t}(i,j) = T^{\tt{t}}_{t}(i,j)-T^{\tt{d}}_{t}(i,j)$
\end{enumerate}

\begin{figure}[!htb]
    \begin{subfigure}[t]{0.27\textwidth}
        \includegraphics[width=1\textwidth]{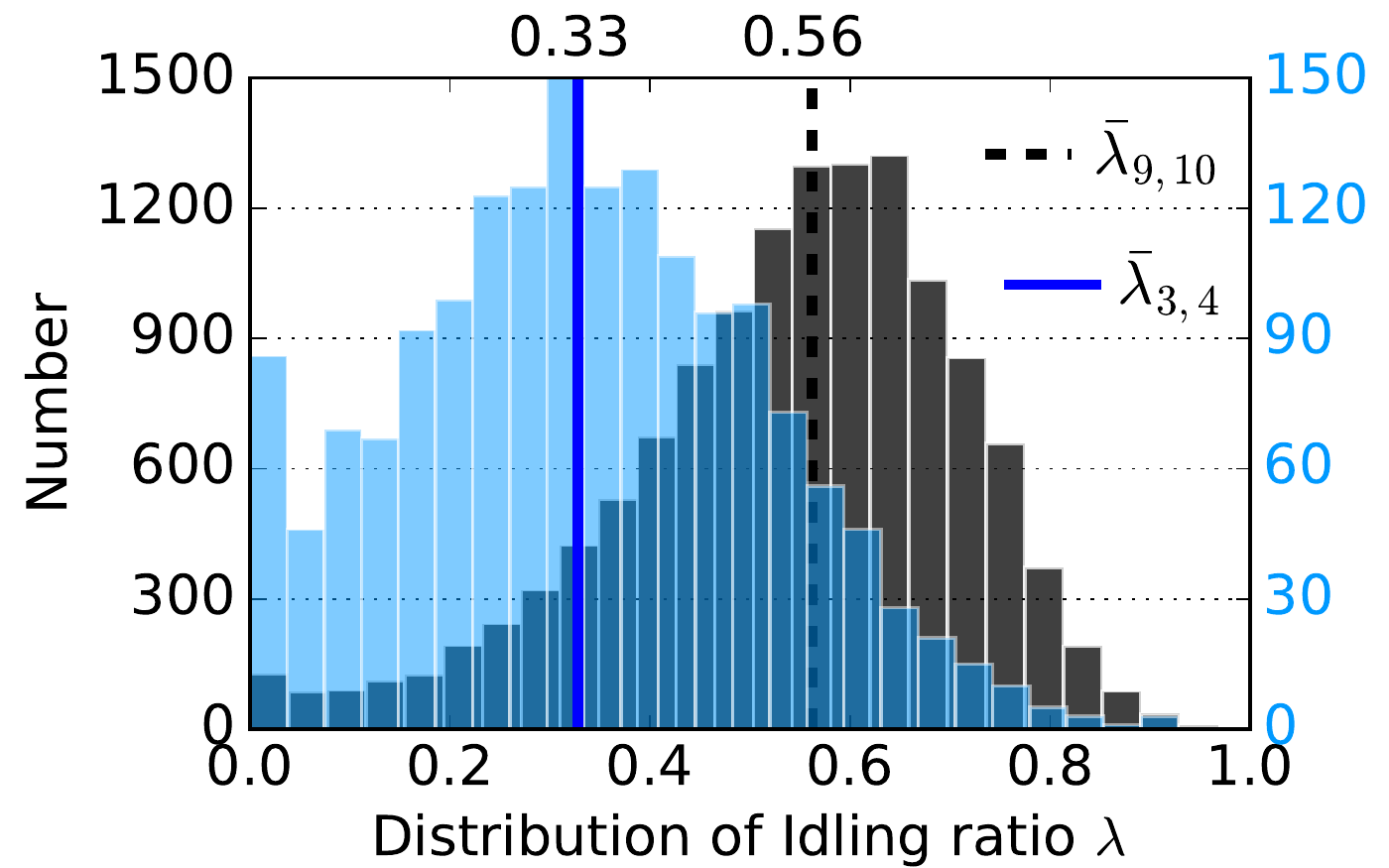}
        \caption{Distribution of idling ratios for 3-4 AM and 9-10 AM}
    \end{subfigure}
    ~
    \begin{subfigure}[t]{0.20\textwidth}
    \includegraphics[width=1.05\textwidth]{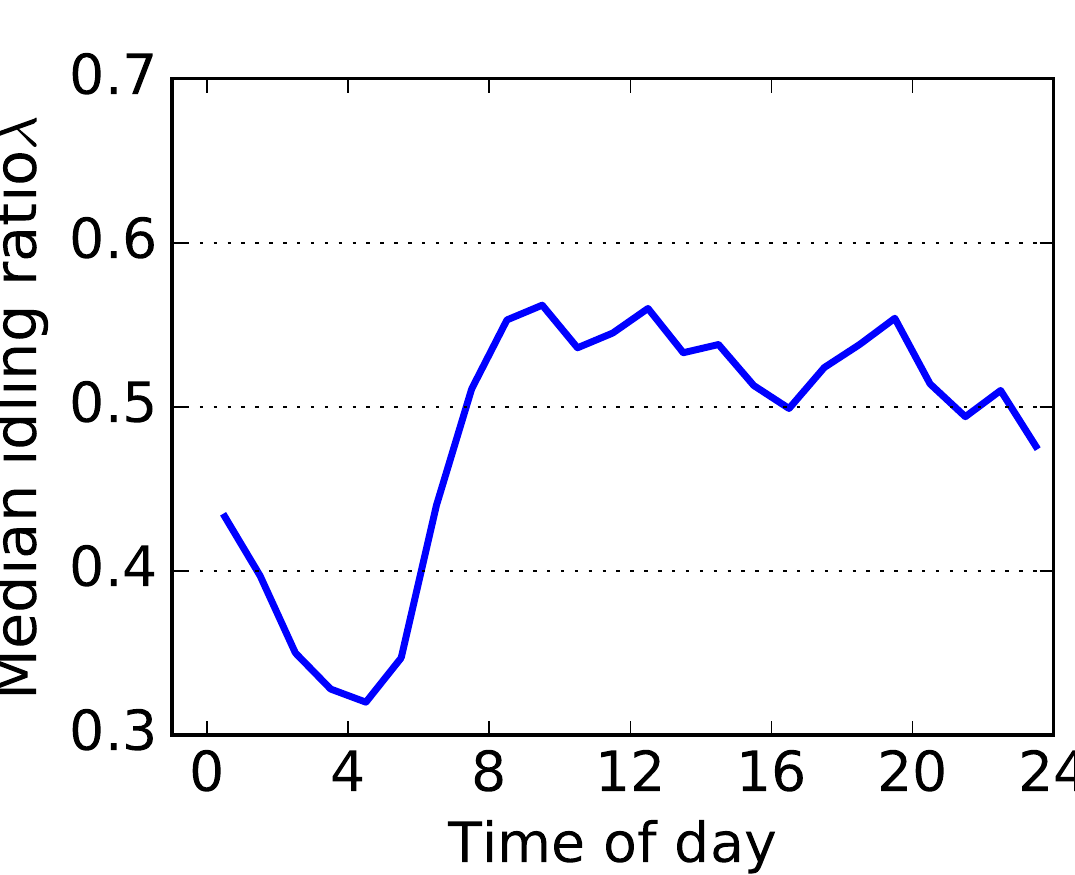}
    \caption{Median idling ratios over a day.}
    \label{fig:timeratio2}
    \end{subfigure}
    \caption{Hourly distribution of idling ratios and median idling ratio over a day.}   \label{fig:timeratio}
\end{figure}

To understand traffic conditions, define a metric called the idling ratio of each source and destination pair by $\lambda \triangleq \frac{T^{\tt{i}}_{t}(i,j)}{T^{\tt{t}}_{t}(i,j)}$. Denote by $\bar{\lambda}_{t_1,t_2}$  the median of idling ratio between time $t_1$ and $t_2$ in the distribution. Fig.~\ref{fig:timeratio} shows the distribution of idling ratios. We observe that the median is 56\% for 9-10 AM, but only 33\% for 3-4 AM due to less traffic.

\subsection{Passenger Pick-up Probability $P^{\tt{p}}_{t}(i)$}

The passenger pick-up probability describes the chance of a taxi driver can pick up a passenger at junction $i$ at time $t$. Following the idea in \cite{rong2016taximarkov}, we use the numbers of taxis and pick-ups around a particular junction to calculate the pick-up probability $P^{\tt{p}}_{t}(i)$ in $\tau$ mins. First, denote the number of pick-ups at junction $i$ from time $t$ to $t+\tau$ by $N^{\tt{p}}_{t:t\plus\tau}(i)$. To estimate the number of taxis around junction $i$ in $\tau$ mins,  denote the number of drop-offs from time $t-\tau$ to $t+\tau$ within $\delta$ kilometers distance from junction $i$ by $N^{\tt{d}}_{t\minus\tau:t\plus\tau}(i)$.  Assuming the taxis are vacant after dropping off the passengers and are roaming immediately around junction $i$ within $\delta$ kilometers in $\tau$ mins. Thus, pick-up probability $P^{\tt{p}}_{t}(i)$ can be estimated by
\begin{equation}
P^{\tt{p}}_{t}(i)=\frac{N^{\tt{p}}_{t:t\plus\tau}(i)}{N^{\tt{p}}_{t:t\plus\tau}(i)+N^{\tt{d}}_{t\minus\tau:t\plus\tau}(i)}
\label{eq:pickpr}
\end{equation}

\begin{figure}[!htb]
    \begin{subfigure}[t]{0.24\textwidth}
        \includegraphics[width=1.05\textwidth]{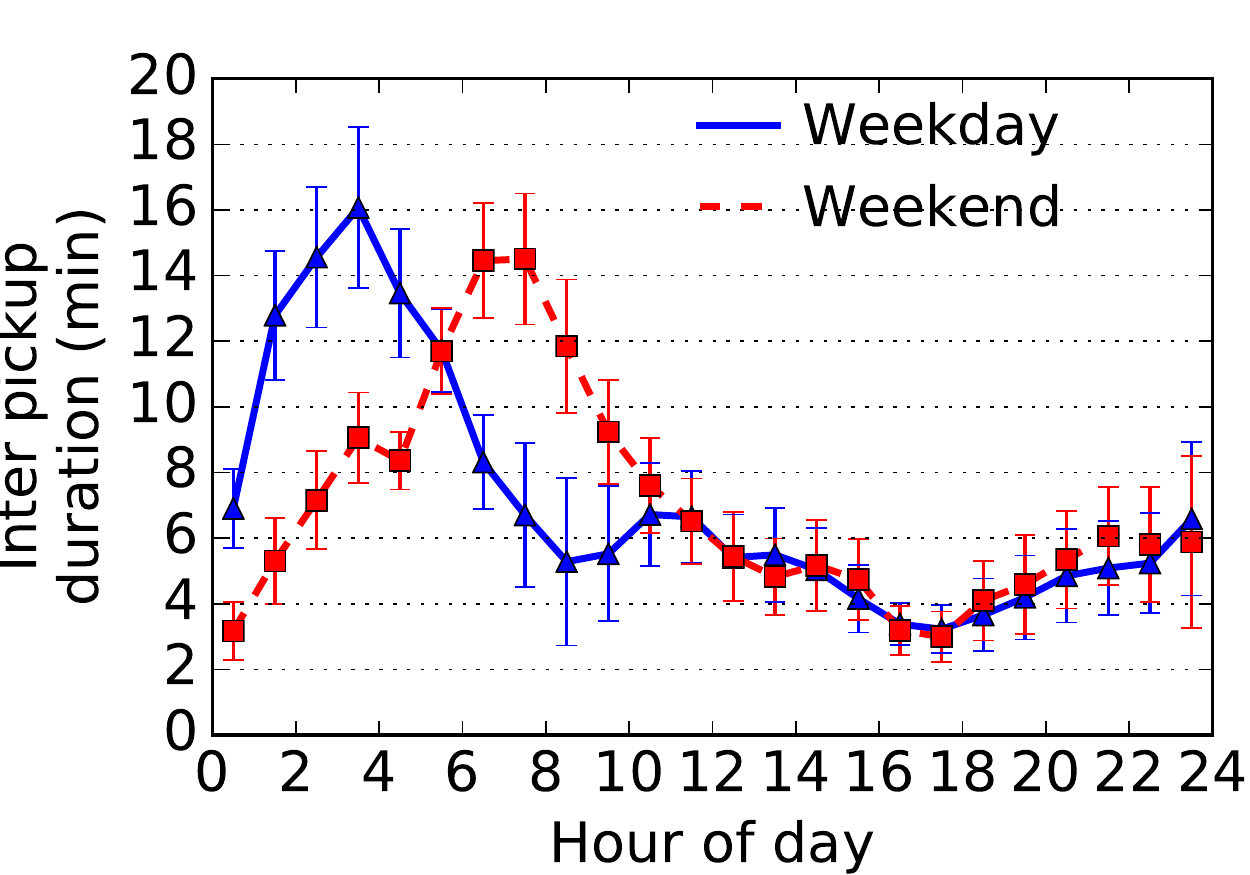}
        \caption{Inter-pick-up durations for weekdays and weekends.}
        \label{fig:pickupduration}
    \end{subfigure}
    \begin{subfigure}[t]{0.24\textwidth}
        \includegraphics[width=1.05\textwidth]{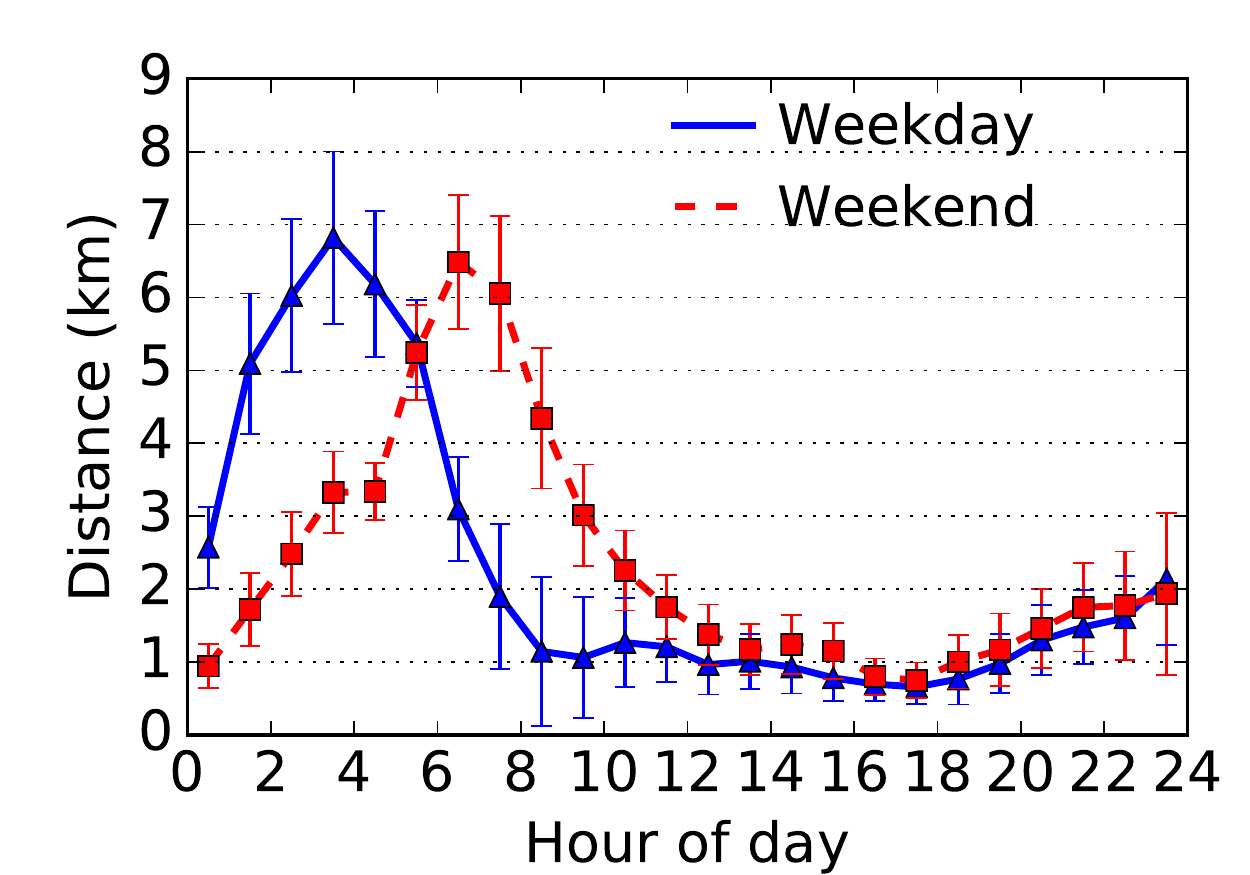}
        \caption{Reachable distances in inter-pick-up durations.}
        \label{fig:pickupdis}
    \end{subfigure}
    \caption{Parameters for estimating pick-up probability.}
\end{figure}

The suitable parameters $\tau$ and $\delta$ can be obtained from the historical taxi trip dataset. For example, $\tau$ can be estimated by the average inter-pick-up duration, the time interval between consecutive pick-ups of a taxi. Using the average driving speed, $\delta$ can be estimated by the reachable distance in the average inter-pick-up duration. Fig.~\ref{fig:pickupduration} depicts the average inter-pick-up durations for weekdays and weekends. We observe that it takes more time to find a passenger at 4 AM on weekday and at 7 AM at weekends. Fig.~\ref{fig:pickupdis} depicts the respective reachable distance in inter-pick-up duration.

In the following study, we set time-varying $\tau$ and $\delta$  according to the average inter-pick-up duration and the respective reachable distance from taxi trip dataset for each hour.

\subsection{Passenger Destination Probability $P^{\tt{d}}_{t}(i,j)$}

The passenger destination probability describes the chance that a passenger needs to commute from one junction to another junction. This probability is time-dependent, because, for example,  passengers are more likely to commute from living places to offices in working hours. One-hour timeslot is used to estimate passenger destination probability from taxi trip dataset. In each timeslot, we obtain the number of trips between each pair of source and destination, and then is normalized by the total number of trips.
Denote the destination probability from junction $i$ to junction $j$ at time $t$ by $P^{\tt{d}}_{t}(i,j)$.  Denote the number of pick-ups at junction $i$ by $N^{\tt{p}}_{t}(i)$, and the number of corresponding drop-offs at junction $j$ by $N^{\tt{d}}_{t}(i, j)$. The passenger destination probability from junction $i$ to junction $x$ is estimated by
\begin{equation}
  P^{\tt{d}}_{t}(i,j) = \frac{N^{\tt{d}}_{t}(i,j)}{N^{\tt{p}}_{t}(i)}
\end{equation}

\subsection{Energy Consumption $E^{\tt{e}}_{t}(i,j)$}

We use a black-box approach to estimate the energy consumption for EVs, based on the work in \cite{cmtseng2017dte,ckt2017phevopt}. The energy consumption model is based on the average driving speed and auxiliary loading. The total energy consumption can be decomposed into moving energy consumption and auxiliary loading energy consumption, which can be estimated by multivariate linear models (see \cite{cmtseng2017dte,ckt2017phevopt} for details):
\begin{align}
E^{\tt{e}}_{t}(i,j) =\ & E^{\tt{mv}}_{t}(i,j)+E^{\tt{ax}}_{t}(i,j) \label{eq:egy}\\
E^{\tt{mv}}_{t}(i,j) =\ & \beta(\alpha_1 v_{t}(i,j)^2+\alpha_2 v_{t}(i,j) +\alpha_3)\cdot D(i,j)\\
E^{\tt{ax}}_{t}(i,j) = \ & \ell_t T^{\tt{t}}_{t}(i,j)/60
\end{align}
where $v_{t}(i,j)$ is the driving speed between junctions $i$ and $j$ at time $t$, obtained from driving speed network. $D(i,j)$ is the driving distance between junction $i$ and junction $j$.

The auxiliary loading $\ell_t$ is highly affected by weather temperatures which is time variant. The auxiliary loading can be estimated from the historic weather temperature and the average auxiliary loading measurements at particular temperatures\footnote{ See \cite{auxload} for an empirical measurement study}. According to New York historical weather and suggested power load, the average auxiliary loading is between 1.5 to 1 kW. The parameter $\beta$ represents aggressiveness factor to capture the driving behavior. Driver behavior has an impact on the energy consumption of vehicles, as driving range will be significantly decreased by aggressive acceleration and deceleration. Mild driving behavior can save up to 30\% to 40\% energy consumption comparing with aggressive driving behavior \cite{bungham2012drivingbehavior,lei2013drivinghev}. Therefore, we define three classes of driving behaviors: i) mild drivers ($\beta=0.8$), ii) normal drivers ($\beta=1$), and iii) aggressive drivers ($\beta=1.2$). Based on previous work \cite{cmtseng2017dte}, the parameters of energy consumption model for Nissan Leaf are set as  $\alpha_1 = 0.1554,  \alpha_2=-5.4634, \alpha_3=189.297$.

\begin{figure*}[!htb]
    \begin{subfigure}[t]{0.26\textwidth}
        \includegraphics[width=\textwidth]{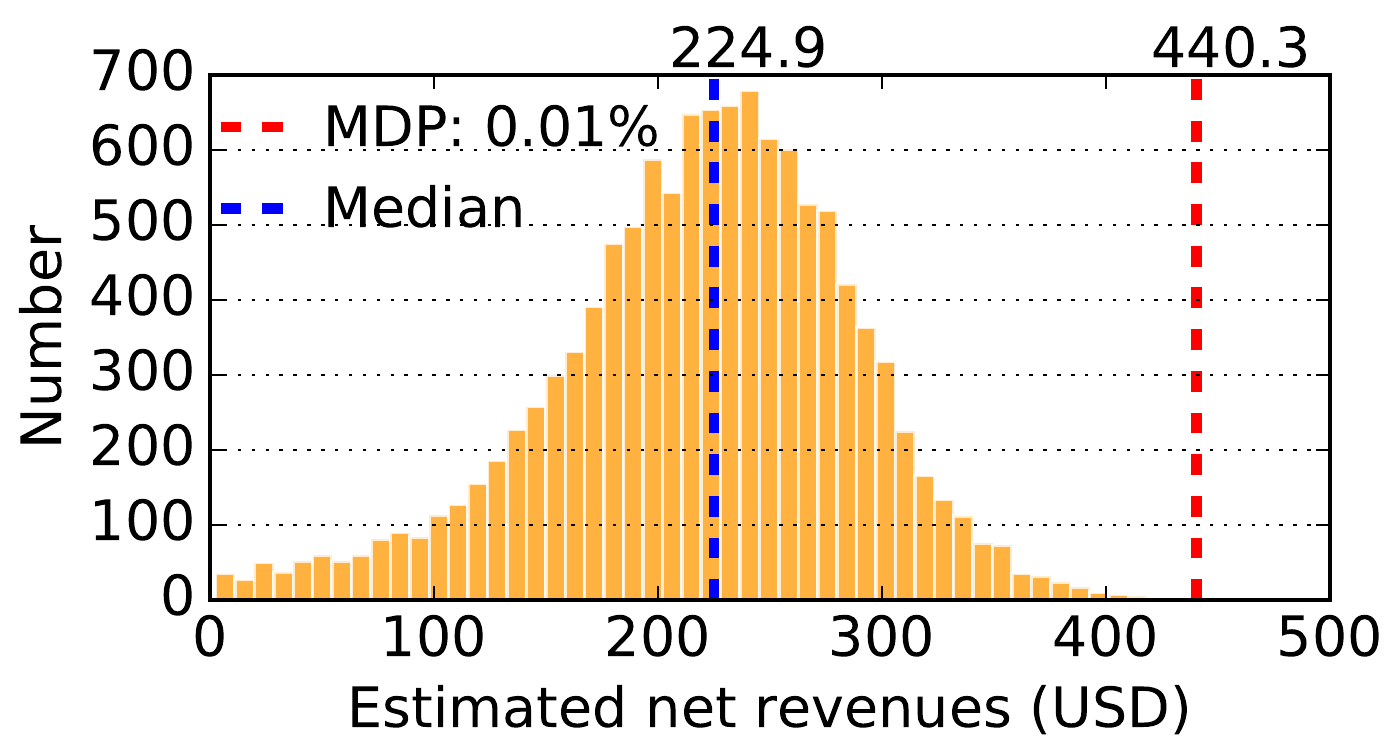} 
        \caption{Estimated net revenues for morning shifts.}
        \label{fig:profitM}
    \end{subfigure}
    \hspace{-0.em}
    \begin{subfigure}[t]{0.23\textwidth}
        \includegraphics[width=\textwidth]{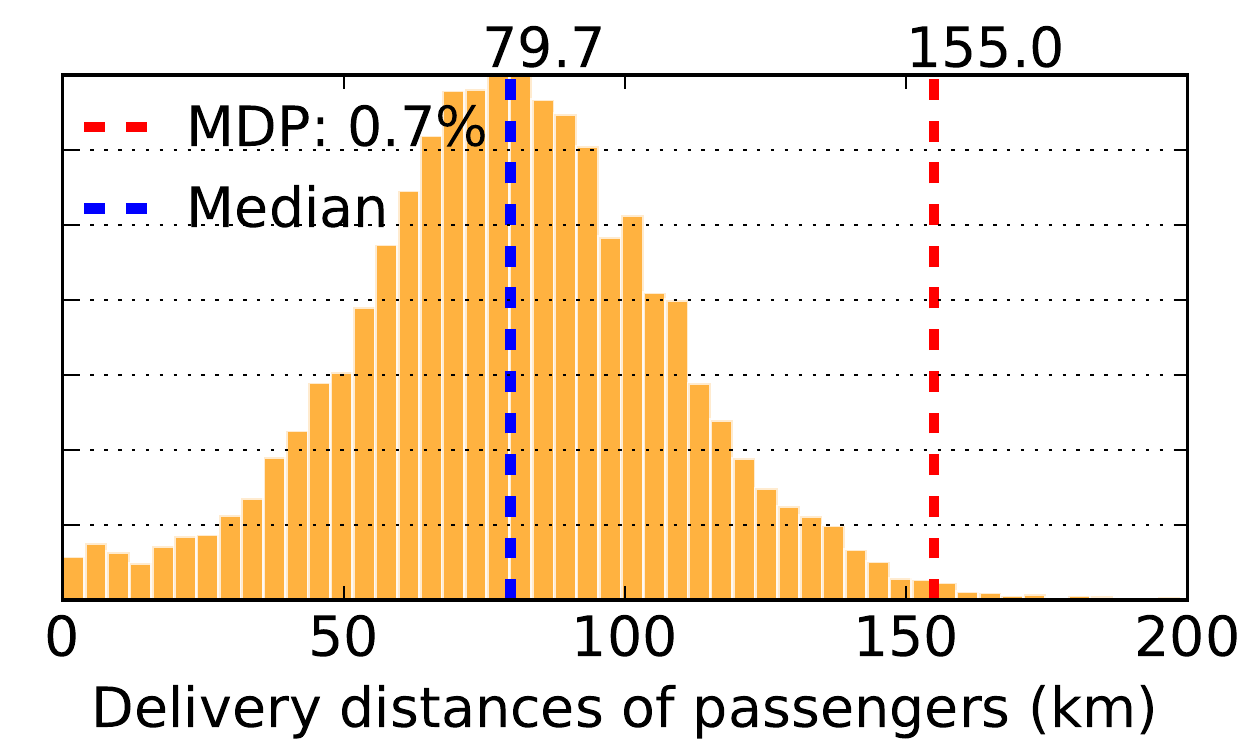}
        \caption{Delivery distances of passengers for morning shifts.}
        \label{fig:disM}
    \end{subfigure}
    \hspace{-0.em}
    \begin{subfigure}[t]{0.23\textwidth}
        \includegraphics[width=\textwidth]{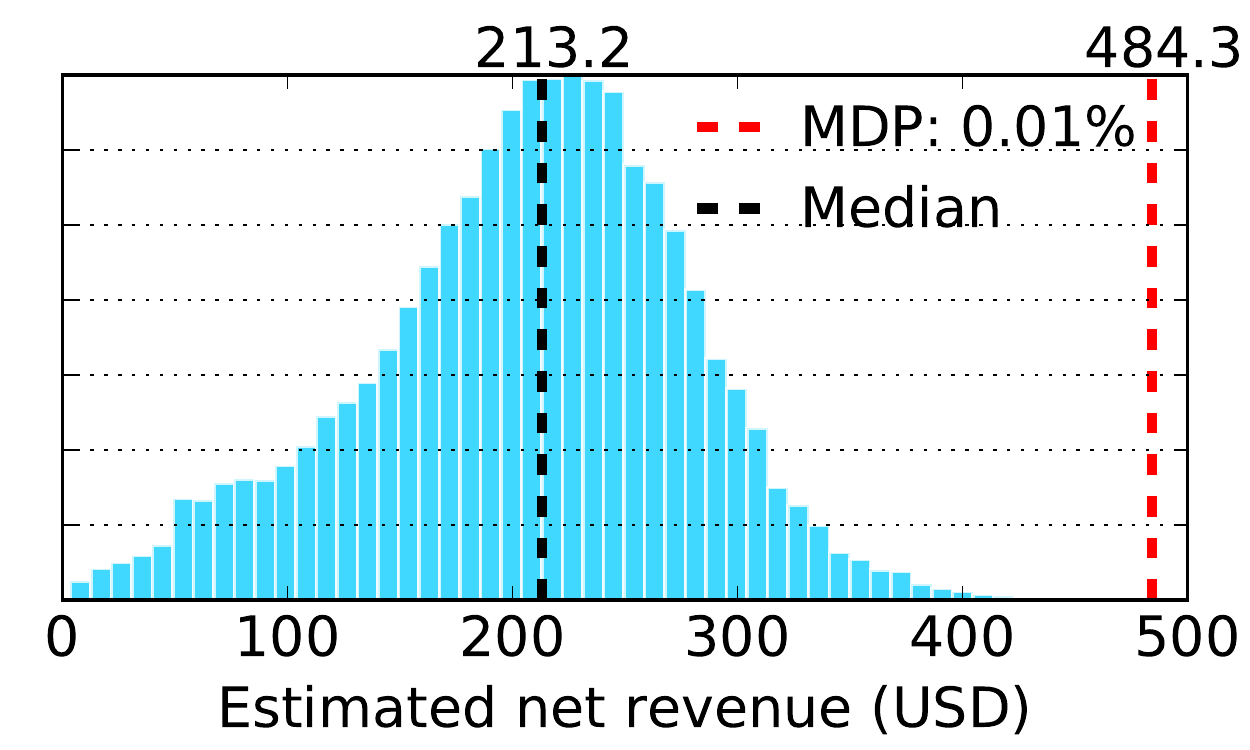}
        \caption{Estimated net revenues for evening shifts.}
        \label{fig:profitN}
    \end{subfigure}
    \hspace{-0.em}
    \begin{subfigure}[t]{0.23\textwidth}
        \includegraphics[width=\textwidth]{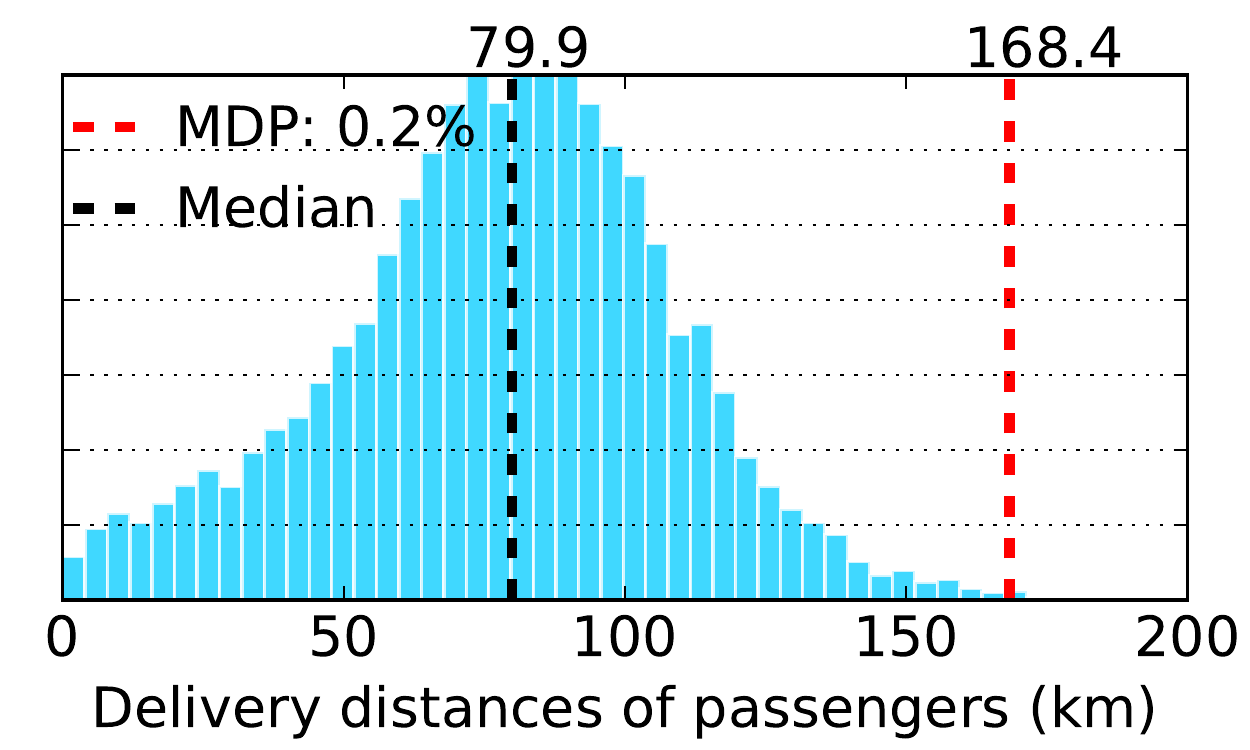}
        \caption{Delivery distances of passengers for evening shifts.}
        \label{fig:disN}
    \end{subfigure}
    \caption{Distributions of estimated net revenues and delivery distances of passengers.}
    \label{fig:ICEdailyanalysis}
\end{figure*}

\begin{figure*}[!htb]
    \begin{subfigure}[t]{0.25\textwidth}
        \includegraphics[width=\textwidth]{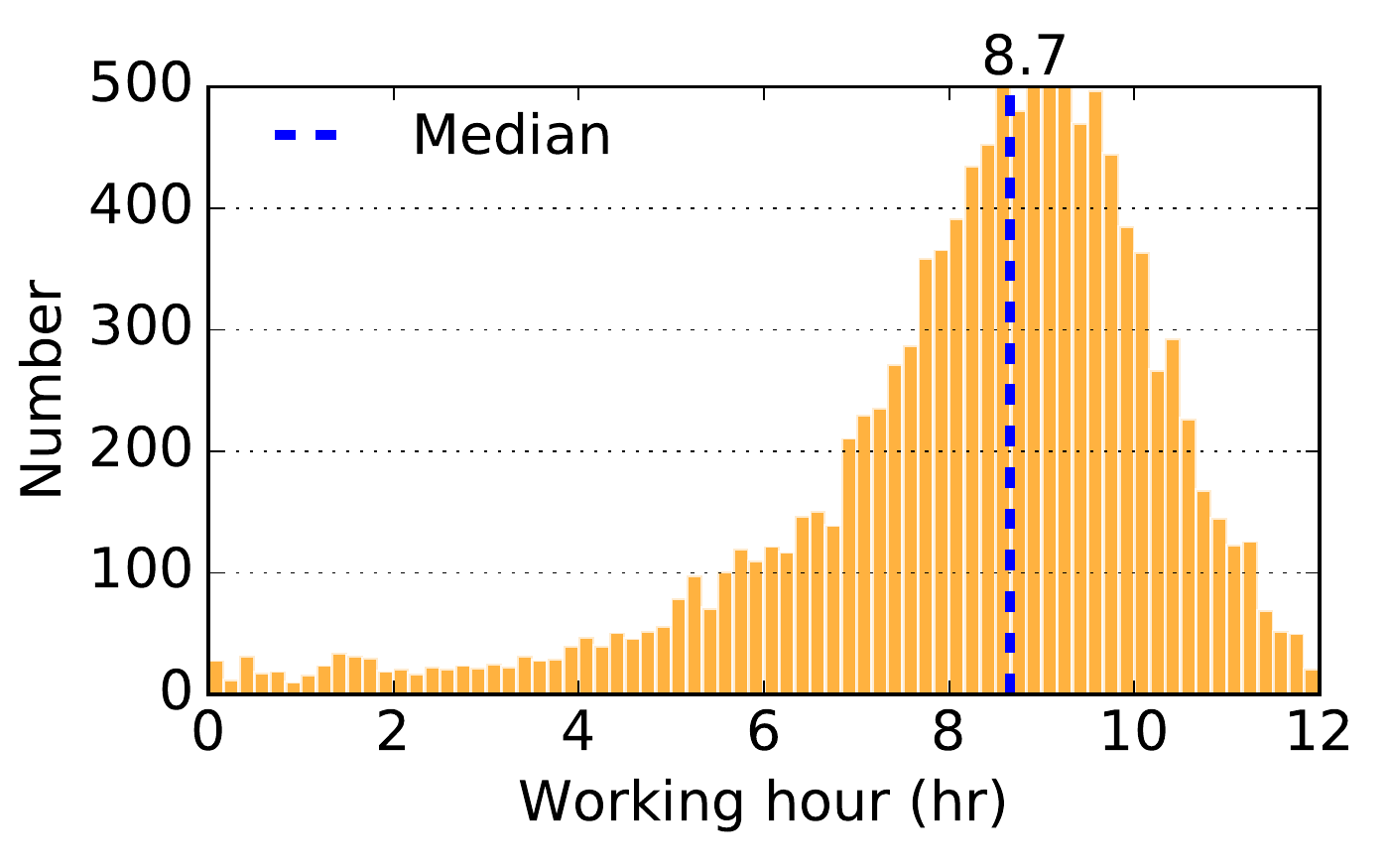} 
        \caption{Working hours for morning shifts.}
        \label{fig:ICEworkhr}
    \end{subfigure}
 \hspace{-0.em}
    \begin{subfigure}[t]{0.225\textwidth}
        \includegraphics[width=\textwidth]{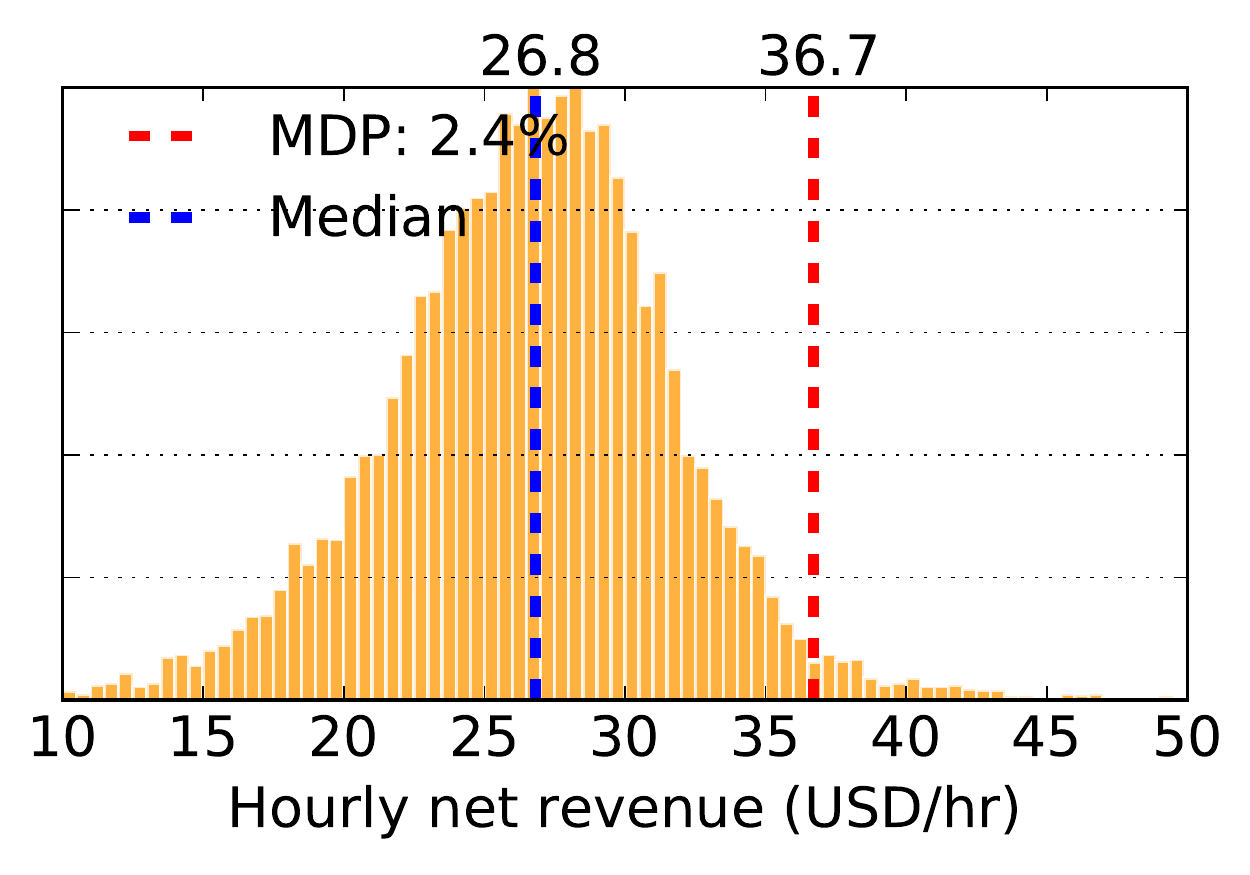} 
        \caption{Estimated hourly net revenues for morning shifts.}
        \label{fig:ICEwage}
    \end{subfigure}
 \hspace{-0.em}	
    \begin{subfigure}[t]{0.25\textwidth}
        \includegraphics[width=\textwidth]{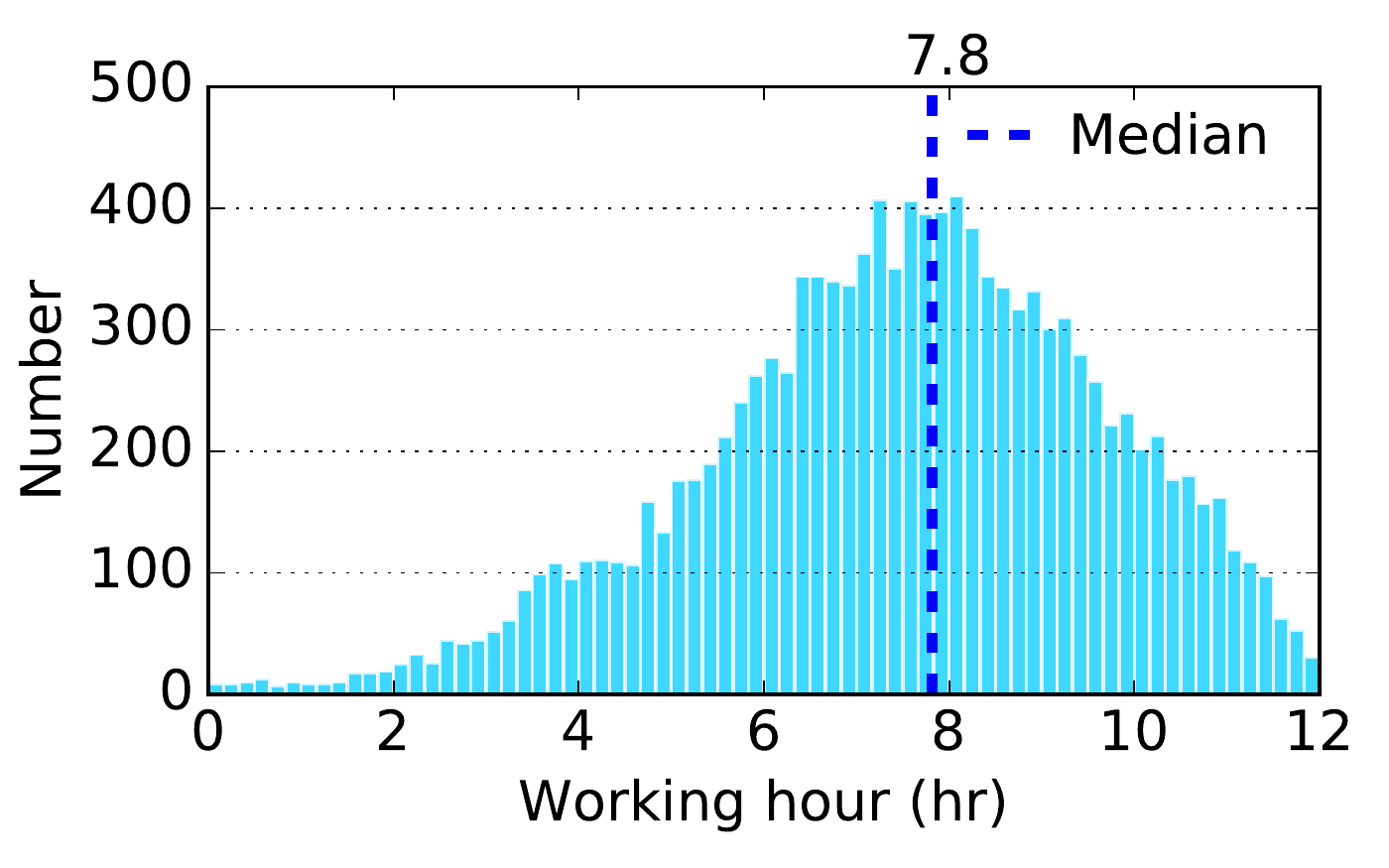} 
        \caption{Working hours for evening shifts.}
        \label{fig:ICEworkhrN}
    \end{subfigure}
 \hspace{-0.em}	
    \begin{subfigure}[t]{0.225\textwidth}
        \includegraphics[width=\textwidth]{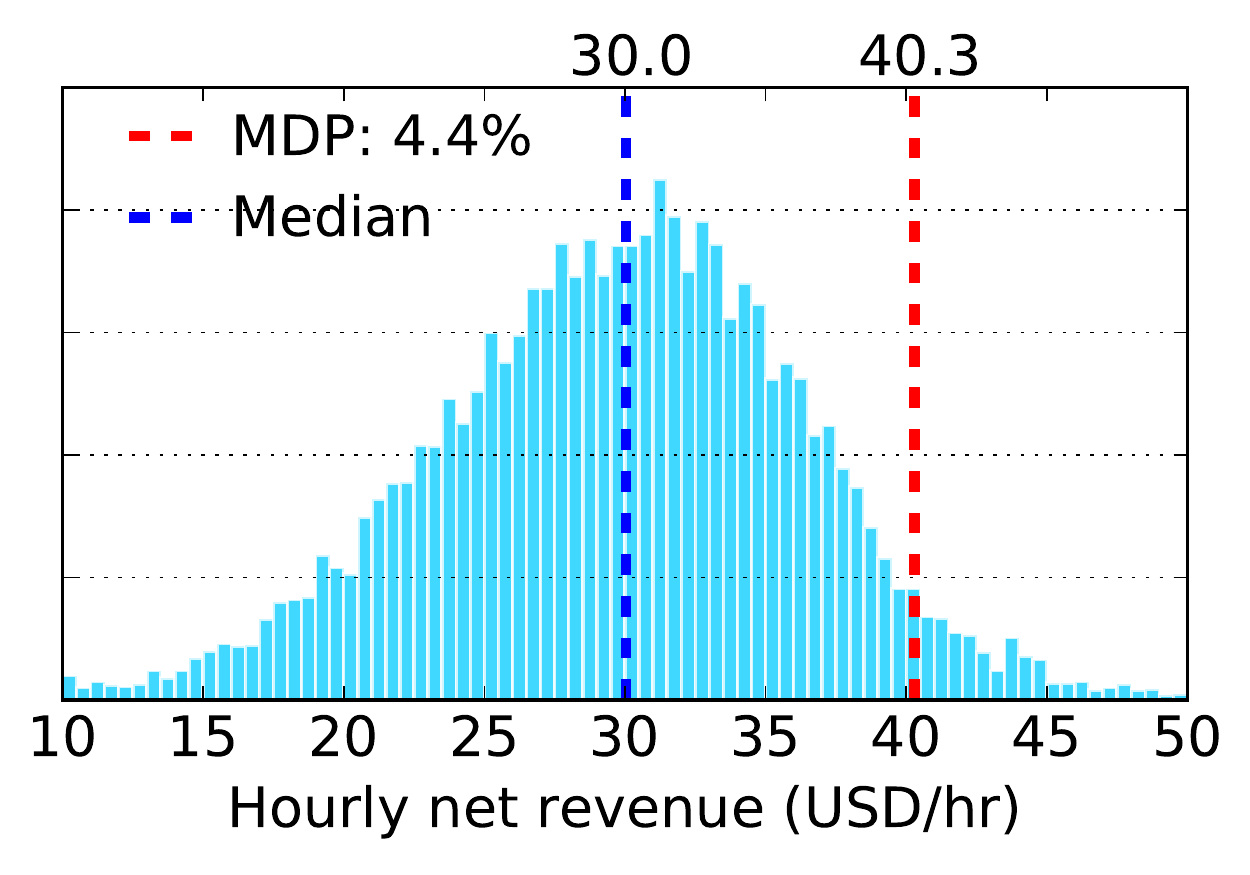} 
        \caption{Estimated hourly net revenues for evening shifts.}
        \label{fig:ICEwageN}
    \end{subfigure}
    \caption{Distributions of working hours and estimated hourly net revenues.}
    \label{fig:wageanalysis}
\end{figure*}

\subsection{Energy Consumption $E^{\tt{e}}_{t}(i,r(i))$}

The electric taxis should arrive at each junction with certain battery state, which can guarantee them to reach the nearest charging stations. The locations of NYC charging station data are obtained from \cite{nychg}. We consider the charging stations for general EVs. Note that there are other charging stations requiring memberships, and are not considered in this study.
To estimate the minimum required energy consumption $E^{\tt{e}}_{t}(i,r(i))$ to the nearest charging station $r(i)$ at junction $i$ at time $t$, the minimum distance between the junction and the nearest charging station is obtained as follows:
\begin{enumerate}
\item Spatialite is used to find the nearest charging station $r(i)$ for junction $i$ in the road network by the shortest distance.
\item The shortest distance is converted into the required driving time based on the driving speed network.
\item The median idling ratio $\bar{\lambda}$ is used to estimate the idling time at time $t$.
\item Given the driving speed network and idling time, the energy consumption $E^{\tt{e}}_{t}(i,r(i))$ is obtained by Eqn.~(\ref{eq:egy}).
\end{enumerate}

\subsection{Taxi Net Revenue $F_{t}(i,j)$}

The fares are calculated according to the rules for New York taxis. Since there are different kinds of surcharge based on times and days, the fare is time-dependent, because of various surcharges\footnote{The initial charge is \$2.50. Plus 50 cents per 1/5 mile or 50 cents per 60 seconds in slow traffic or when the taxi is stopped. 50-cent MTA State Surcharge is required for all trips that end in New York City. Another 30-cent Improvement Surcharge is required.  Daily 50-cent surcharge is required from 8pm to 6am. \$1 surcharge is required from 4pm to 8pm on weekdays, excluding holidays. Toll fees are ignored since the taxi driver will not receive any revenue from tolls. }.
The net revenue of a trip can be calculated by deducting fuel/electricity cost from the revenue. Therefore, the net revenue of a trip from junction $i$ to junction $j$ at time $t$ is
\begin{equation}
F_{t}(i,j) = F^{\tt{R}}_{t}(i,j) - E^{\tt{e}}_{t}(i,j) \cdot U
\end{equation}
where $F^{\tt{R}}_{t}(i,j)$ is the recorded amount of base fare plus the surcharges from $i$ to $j$ at time $t$, and $U$ is the unit price.

\subsection{Charging Rate $C$}

Two types of charging rates are considered in this study: mode 3 charging and (direct current)  fast charging. Currently, mode 3 charging is more common than fast charging. The charging power of mode 3 charging is 6.6 kW (e.g., for Nissan Leaf), whereas the charge power of fast charging is 50 kW.

\section{Evaluation based on NYC Taxi Trip Dataset}

In this section, we apply the MDP to optimize computerized taxi service strategies and evaluate the improvement in net revenues using NYC  taxi trip dataset. We first examine the net revenue of conventional ICE taxis and improvement by MDP under a basic setting with complete knowledge of taxi trip information for one single taxi, which represents the best-case scenario. Next, we study a similar setting for electric taxis. Then, we relax the basic setting by more realistic settings: (1) using only historical data as training dataset,  (2) an extension to multiple taxis, and (3) considering different driving behavior.

\subsection{Basic Setting of ICE Taxi}

{\bf Setting:}  This section presents an evaluation study based on one-day data of January 9 2013 in the NYC taxi trip dataset. In Sec.~\ref{sec:emi}, an evaluation using a whole year's data will be presented.  First, we note that the NYC  taxi trip dataset has only records of trip distance and duration, and pick-up and drop-off information. There is no full mobility data trace of taxis, in particular when the taxis are roaming without passengers. It is difficult to estimate the exact total travel distance (i.e., including roaming and passenger delivery). Hence, we estimate a lower bound for the total travel distance by connecting the shortest path between a drop-off location and a subsequent pick-up location. As such, we obtain an optimistic estimation of net revenue (i.e., revenue minus fuel cost) by the lower bound of total travel distance.
We consider a basic setting, such that the optimal policy of MDP is employed in one single taxi, based on complete knowledge of taxi trip information on the same day from the dataset. In  Sec.~\ref{sec:multit}, using historical data for prediction and multiple taxis will be presented.

Note that it is challenging to evaluate the exact performance of modified taxi behavior using historical dataset. For example, when a passenger is picked up by a taxi with modified behavior, who was originally picked up  by another taxi in the dataset, it is not clear how original taxi should behave in the evaluation. Therefore, we consider a simple approach of evaluation, such that other taxis always follow the recorded trajectories as in the dataset, no matter picking up the supposed passengers of the dataset or not. Although this will not attain absolute accuracy, this is a simple approach without the knowledge of the disrupted behavior of other taxis in real life. Note that if we only modify the behavior of a small number of taxis, then this simple approach will give rather accurate evaluation.

Also, refueling is not considered for ICE taxis, because ICE taxi drivers normally fill up the gas tanks between the shifts\footnote{Most NYC taxis operate in two shifts per day. Each normally lasts for 12 hours. More than 40\% of taxi drivers change shifts at around 5 AM or 5 PM. In this study, we assume that a morning shift is from 5 AM to 5 PM, whereas an evening shift is from 5 PM to 5 AM.}. Then the MDP model for an ICE  taxi is identical to that of an electric taxi in Sec.~\ref{sec:mdp}, but without recharging decisions.

{\bf Observations:}  Based on the NYC taxi trip dataset, Fig.~\ref{fig:profitM} shows the distribution of (optimistically) estimated net revenues from all the trips (with 11746 taxi drivers) for morning shifts. The blue dashed line indicates the median of taxi drivers. We observe that 50\% drivers earn above USD\$223. The red dashed line indicates the expected estimated net revenues when a taxi driver follows the optimal policy of MDP assuming 12 working hours. This taxi driver is expected to earn USD\$440.  Therefore, optimizing the taxi service strategy enables a taxi driver to earn at most among the top 0.01\%. Fig.~\ref{fig:disM} shows the delivery distances of passengers per taxi drivers. More than 50\% taxis travel more than 79 kilometers for passenger delivery. By optimizing taxi service strategy, a taxi driver is expected to travel up to 155 kilometers for passenger delivery.
Fig.~\ref{fig:profitN}-\ref{fig:disN} show the distributions for evening shifts. The median net revenue is smaller than that of morning shifts because of shorter working hours (Fig.~\ref{fig:ICEworkhrN}). Also, we observe that the median delivery distances of passengers for the evening shift is similar to that of morning shifts.

The computation of expected net revenue of the optimal policy of MDP assumes 12 working hours. The distribution of working hours for morning shifts is shown in Fig.~\ref{fig:ICEworkhr}. We observe that most of drivers work less than 12 hours, and the median working hours on the day is 8.7 hours.  For a normalized comparison, we also study the hourly net revenues, instead of net revenues per shifts. The distribution of estimated hourly net revenues for morning shifts is presented in Fig.~\ref{fig:ICEwage}. We observe that the hourly net revenue of MDP driver is the top 5\% in both shifts.
We notice that higher hourly net revenue is due to shorter working hour with long trips. 
Fig.~\ref{fig:ICEworkhrN} shows that taxi drivers have shorter working hours for evening shifts, but their hourly net revenues, because of extra surcharge for evening shifts.

\begin{figure*}[!htb]
    \begin{subfigure}[t]{0.23\textwidth}
        \includegraphics[width=\textwidth]{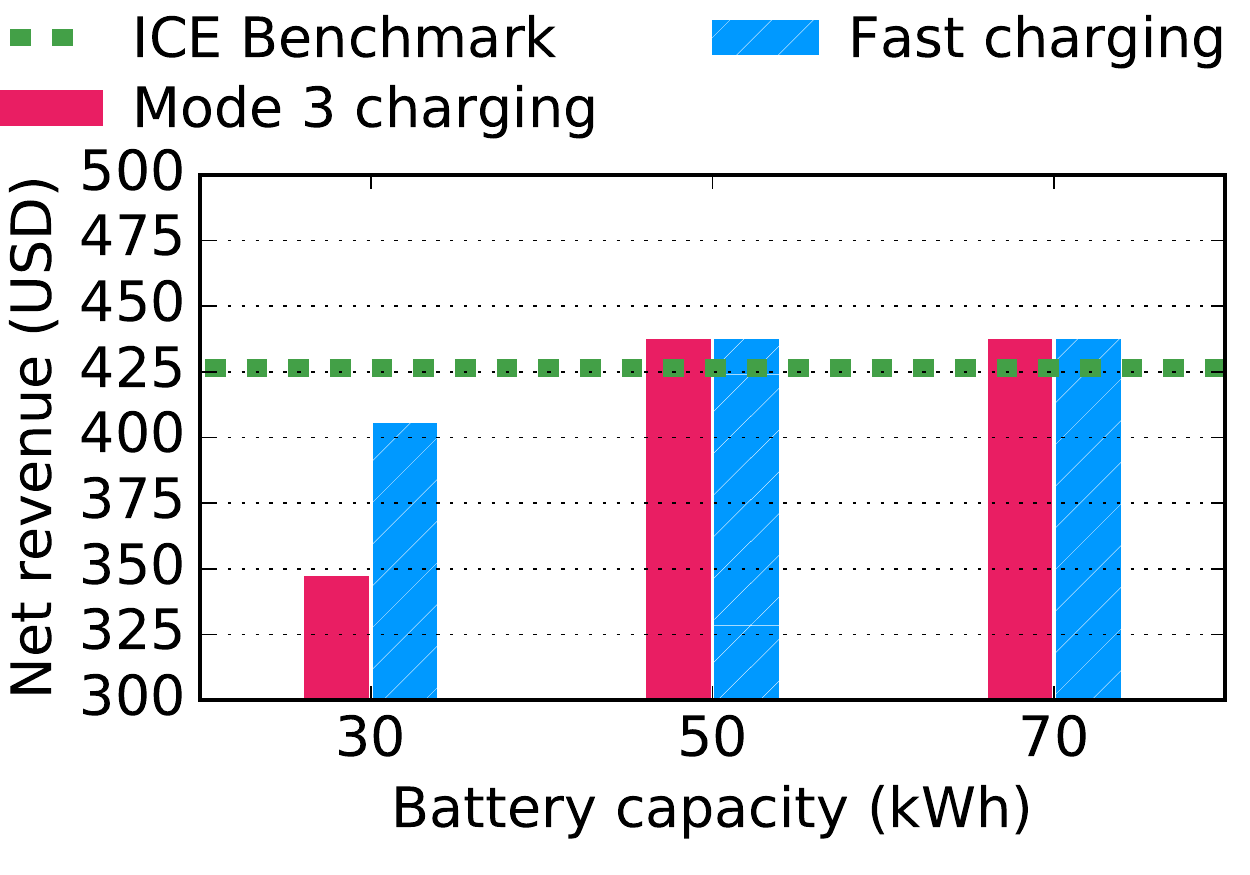} 
        \caption{Estimated net revenues of electric taxis for morning shifts.}
        \label{fig:EVprofit}
    \end{subfigure}
 \hspace{-0.em}
    \begin{subfigure}[t]{0.23\textwidth}
        \includegraphics[width=\textwidth]{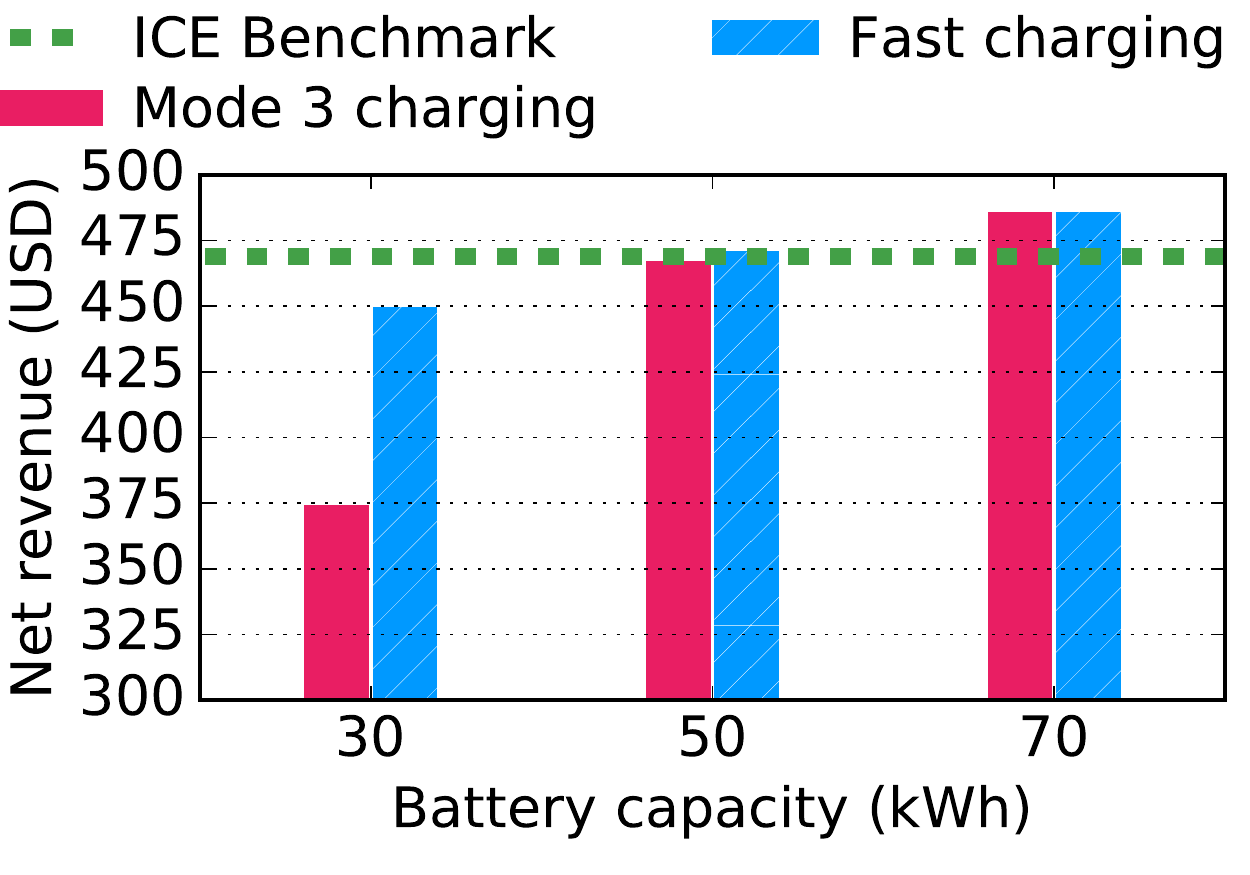} 
        \caption{Estimated net revenues of electric taxis for evening shifts.}
        \label{fig:EVprofitnight}
    \end{subfigure}
 \hspace{-0.em}	
    \begin{subfigure}[t]{0.25\textwidth}
        \includegraphics[width=0.92\textwidth]{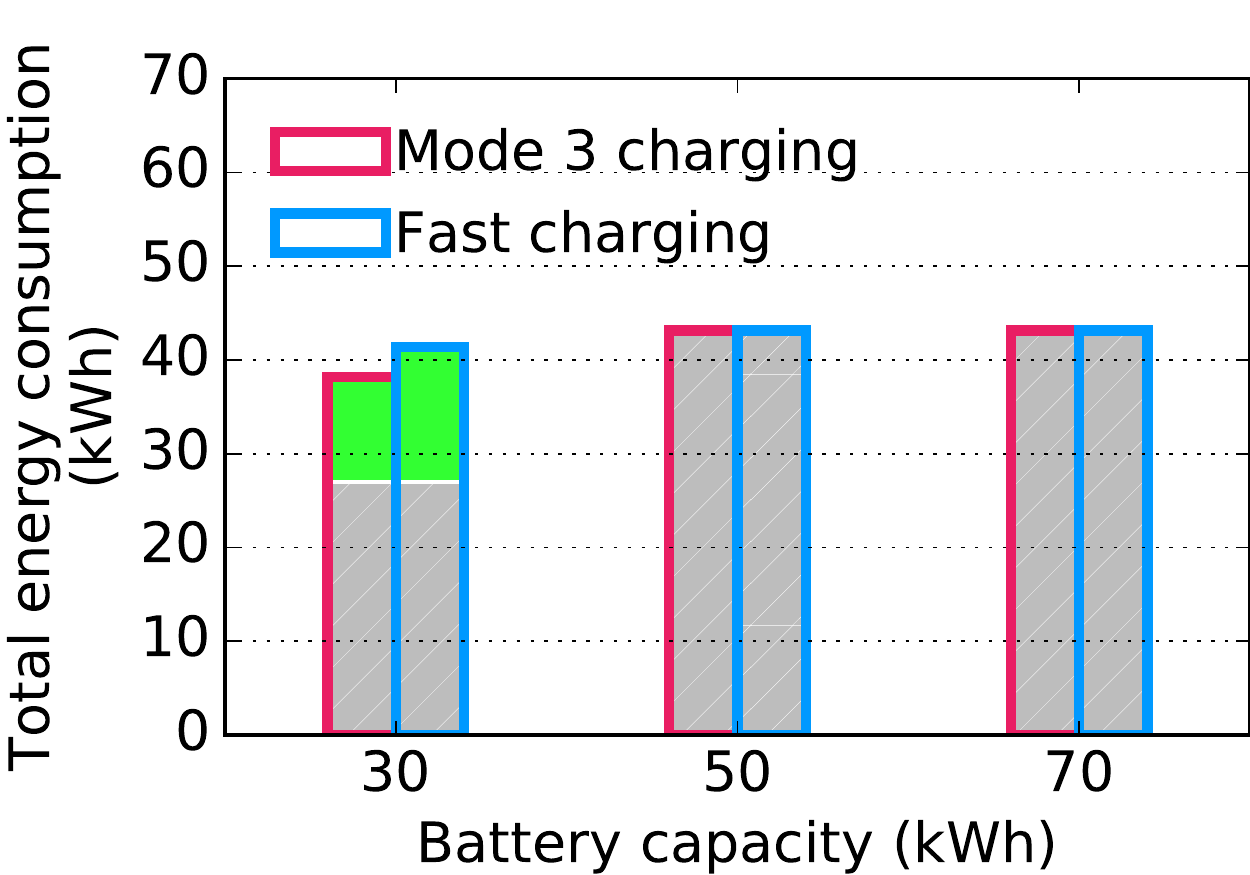} 
        \caption{Energy consumptions for morning shifts.}
        \label{fig:EVdis}
    \end{subfigure}
 \hspace{-0.em}	
    \begin{subfigure}[t]{0.23\textwidth}
        \includegraphics[width=\textwidth]{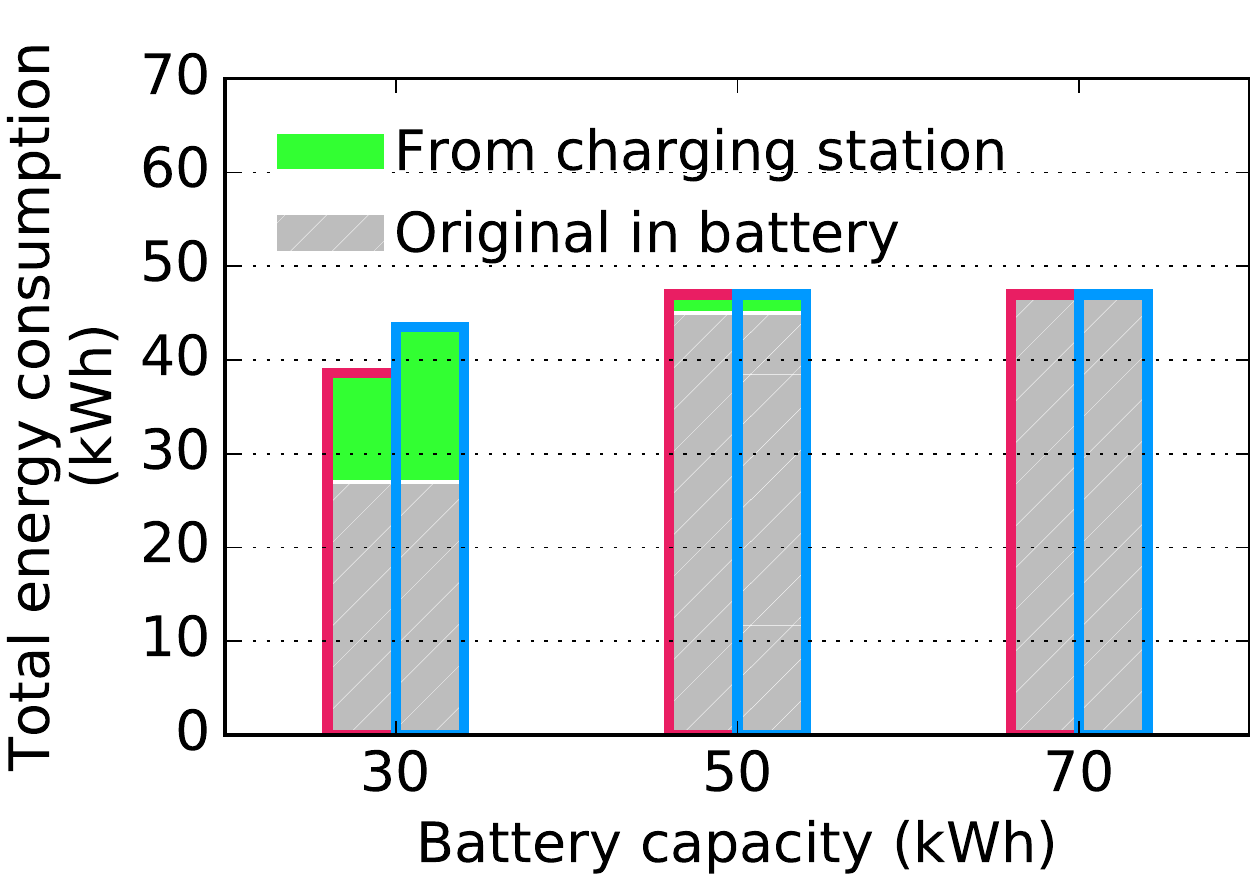} 
        \caption{Energy consumptions for evening shifts.}
        \label{fig:EVdisnight}
    \end{subfigure}
    \caption{Estimated net revenues and energy consumptions of electric taxis.}
    \label{fig:EVanalysis}
\end{figure*}

\begin{figure*}[!htb]
    \begin{subfigure}[t]{0.35\textwidth}
        \includegraphics[width=\textwidth]{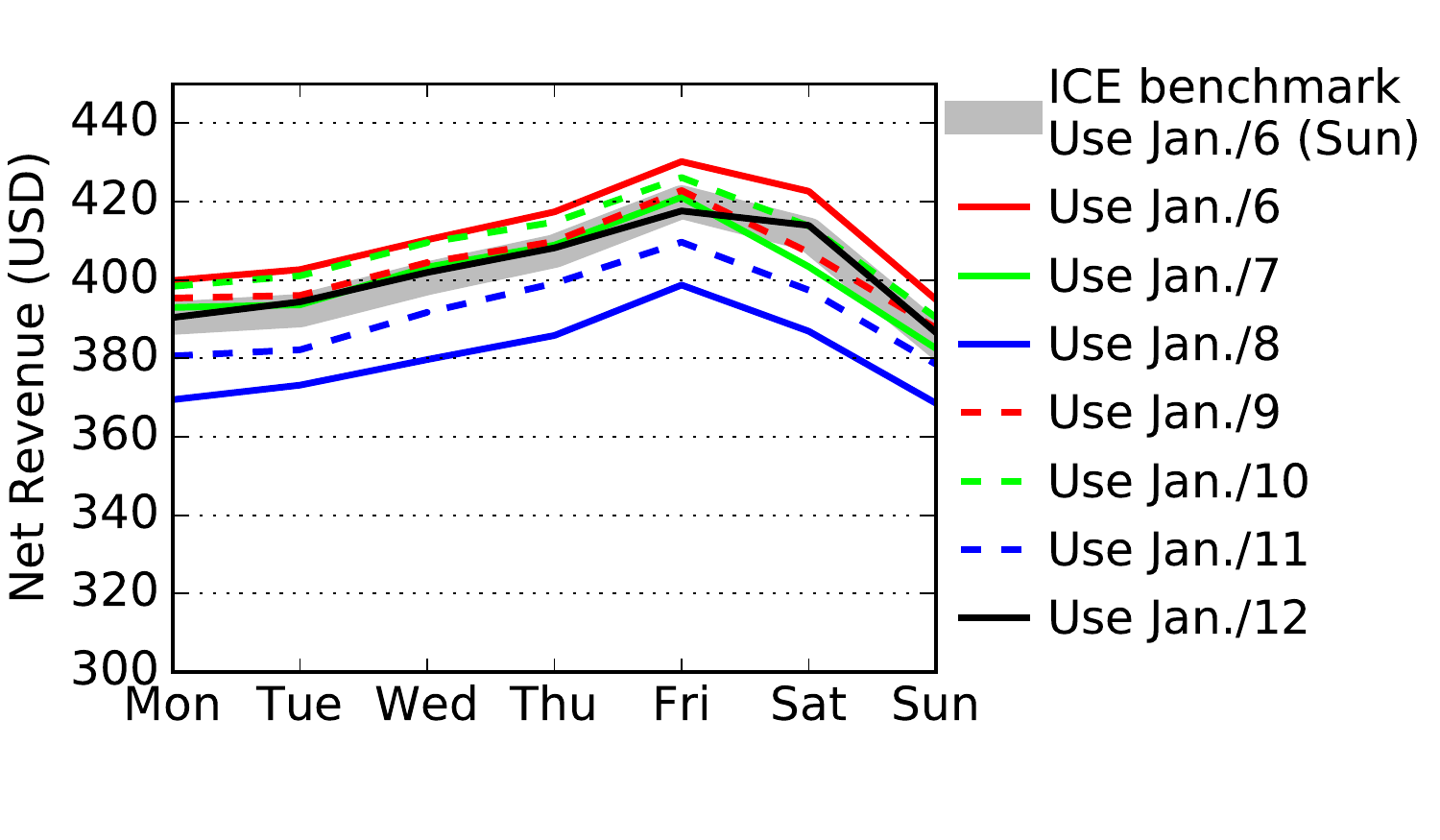} 
        \caption{Estimated net revenues different dates in January as training data with 70 kWh battery capacity.}
        \label{fig:recmd1}
    \end{subfigure}
    ~
    \begin{subfigure}[t]{0.35\textwidth}
        \includegraphics[width=\textwidth]{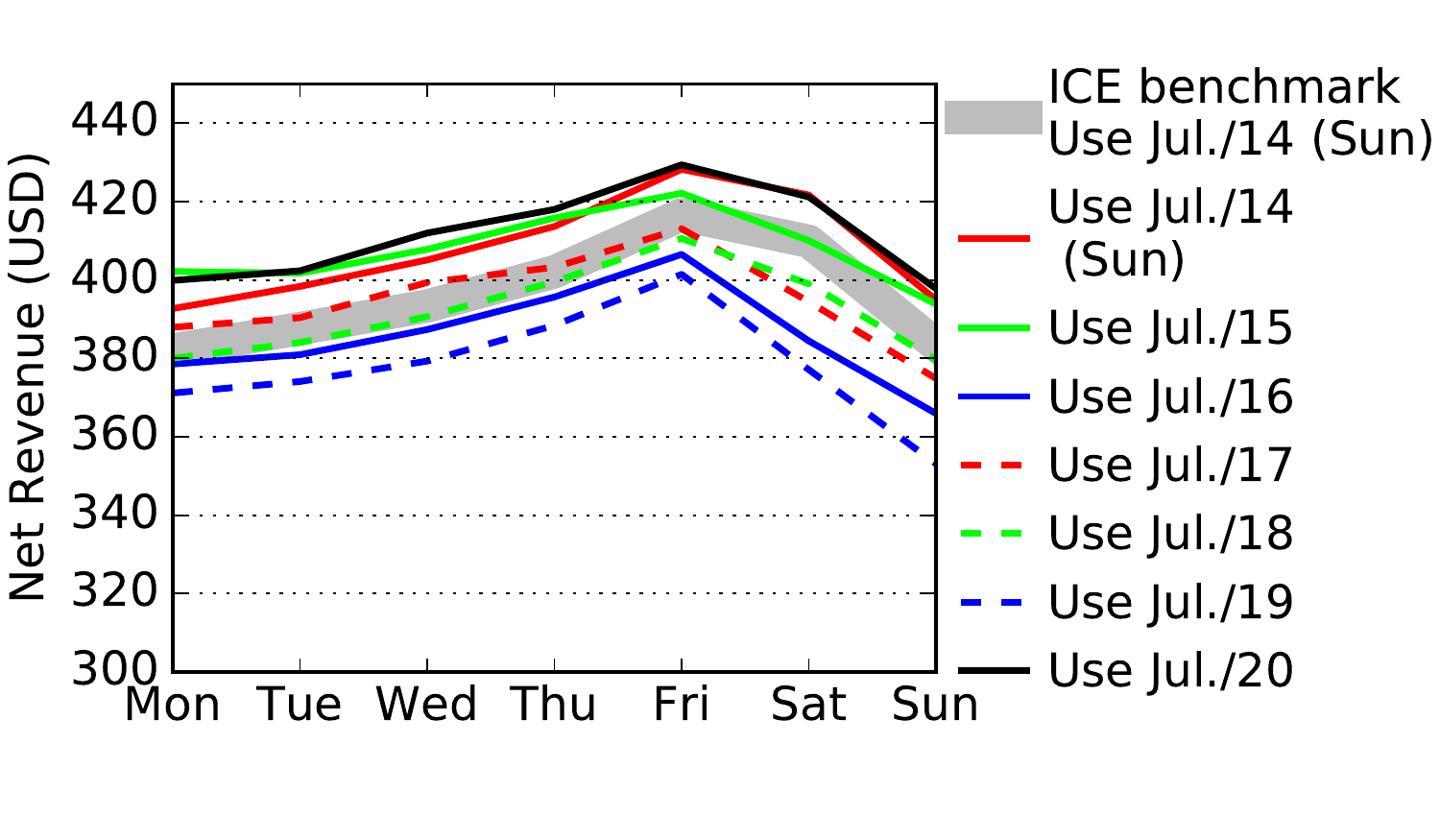} 
        \caption{Estimated net revenues different dates in July as training data with 70 kWh battery capacity.}
        \label{fig:recmd2}
    \end{subfigure}
    ~
    \begin{subfigure}[t]{0.235\textwidth}
        \includegraphics[width=\textwidth]{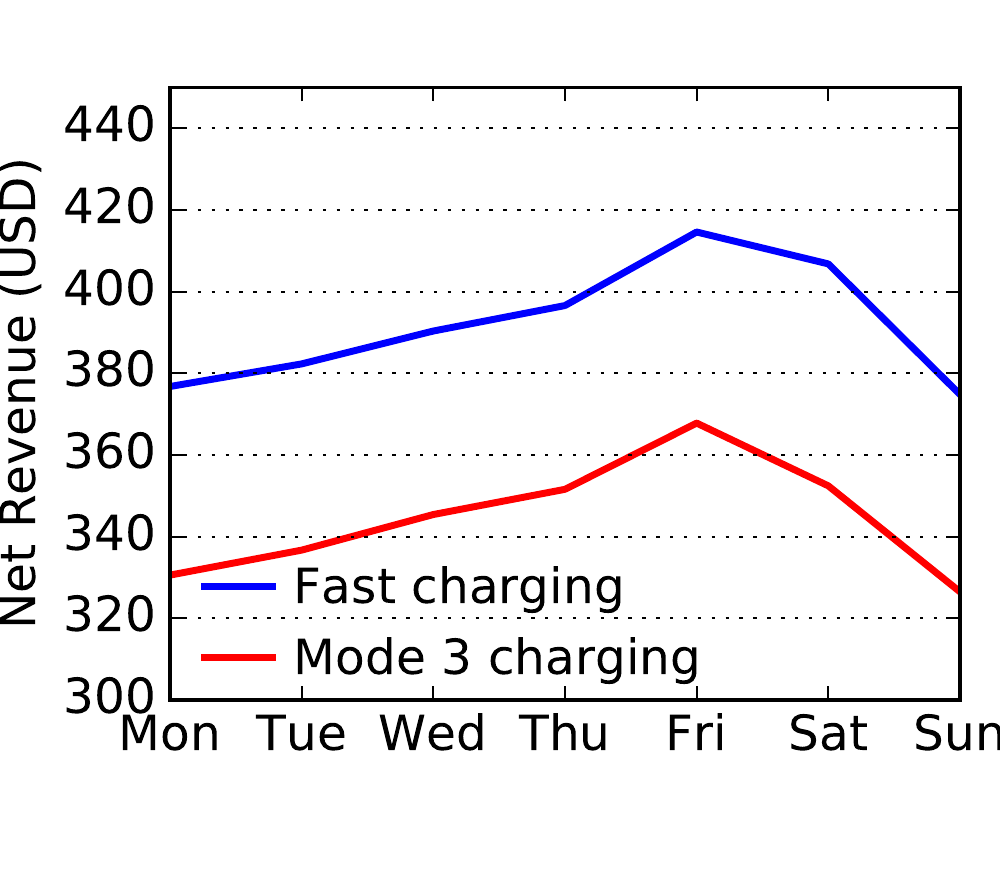} 
        \caption{Estimated net revenues using historical data with 30 kWh battery capacity.}
        \label{fig:recmd3}
    \end{subfigure}
    \caption{Estimated net revenues using historical data for electric taxis.}
    \label{fig:recmd}
\end{figure*}

{\bf Ramifications:} Optimizing taxi service strategies can significantly improve the profitability of taxi drivers. Our evaluation based on a basic setting shows that optimized service strategy for a conventional ICE taxi can earn at most among the top 0.1\%. Although this represents the best-case evaluation, the subsequent sections will relax to more realistic settings, and yet still show a considerable advantage.

\subsection{Basic Setting of Electric Taxi}

{\bf Setting:} In this section, we apply a similar basic evaluation based on the data of January 9 2013 to electric taxis. We employ the energy consumption model of Nissan Leaf \cite{cmtseng2017dte}. In fact, the most determining factor of performance of EVs is battery capacity. Hence,  the energy consumption model of Nissan Leaf suffices to provide a generic estimation of energy consumption of electric taxis. Usually, the EVs will not allowed to be overly re/discharged to protect the battery. Therefore, we set the available battery level from 5\% to 95\% of the capacity. We consider typical settings of battery capacity for EVs (e.g., 30 kWh, 50 kWh, 70 kWh). Each setting can affect the recharging decisions and net revenues considerably.

{\bf Observations:} Figs.~\ref{fig:EVprofit}-\ref{fig:EVprofitnight} show the estimated net revenues for electric taxis under different battery capacities. The blue bars represent the net revenues using fast charging, while the red bars represent those using mode 3 charging. We observe that electric taxis equipped with 50 kWh battery can make comparable net revenues with traditional ICE taxis using fast charging for morning shifts. Note that in general smaller batteries require more frequent recharging, which can reduce revenue. EVs with  smaller batteries are cheaper. The net revenue gap between using fast charging and mode 3 charging is smaller when the battery capacity increases.
The estimated net revenue reaches USD\$438 when battery capacity is above 50 kWh. The net revenue is higher than that of ICE taxis using optimized service strategies (i.e., USD\$426 benchmark), because electricity cost is cheaper.

Figs.~\ref{fig:EVdis}-\ref{fig:EVdisnight} show the driving distances and energy consumptions under different battery capacities. The blue bordered bar represent the driving distances using fast charging while red bordered bars represent  those using mode 3 charging. The green portions represent the amount of charging energy received from charging stations, while gray portions represent the amount from initial batteries.  We observe that the total driving distance is around 242 kilometers without recharging for morning shifts. At  night, the electric taxis are expected to drive longer distances because of less traffic. The required energy consumption without charging for morning shifts is 43 kWh, which can be provided by 50 kWh battery (i.e., 45 kWh usable capacity) without recharging. For evening shifts, the required energy consumption increases to 45.1 kWh. Therefore, electric taxis with 50 kWh battery are then required to recharge during shifts.

{\bf Ramifications:} Optimizing taxi service strategies for electric taxis can improve the profitability of taxi drivers. But the effect depends on the battery capacity. With more capacity (e.g., 50 kWh, 70kWh), the taxi driver can earn comparable net revenue with the one of ICE taxi using optimized service strategy. It is because that recharging will incur inefficiency for electric taxis with a low capacity battery.

\begin{figure*}[!htb]
    \begin{subfigure}[t]{0.2\textwidth}
        \includegraphics[width=\textwidth]{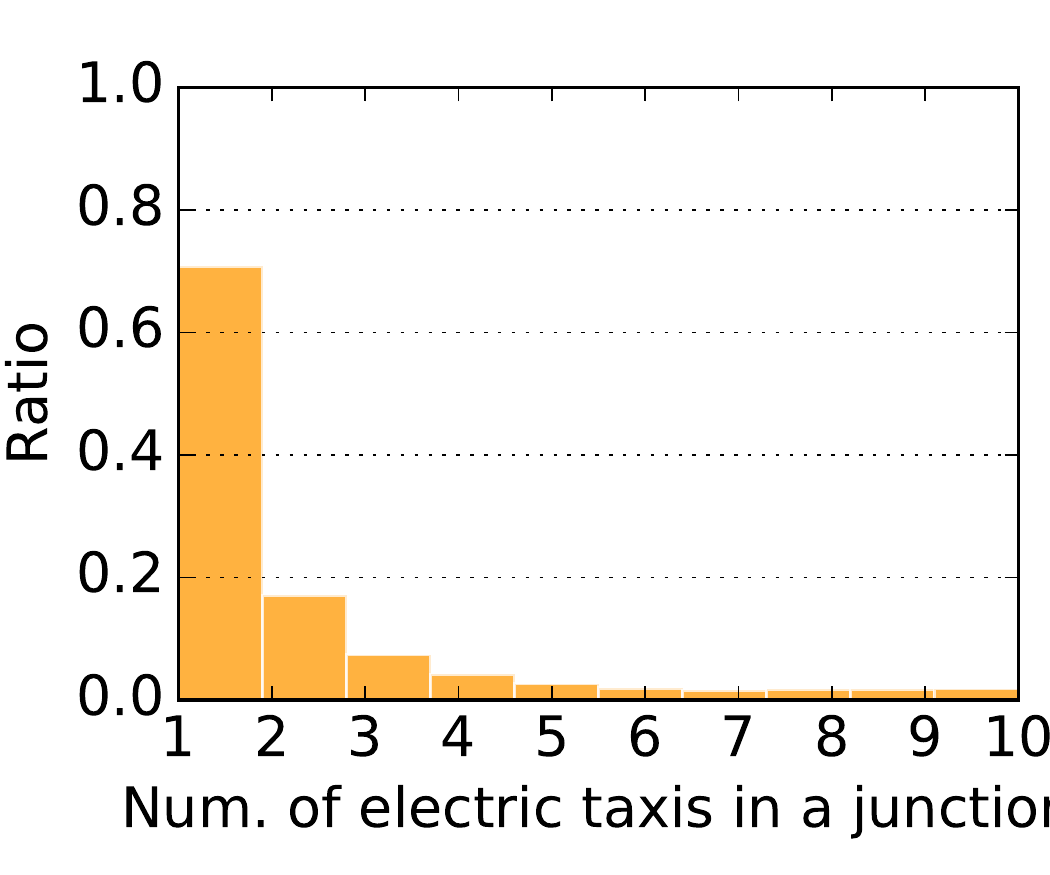} 
        \caption{Histogram of number of taxis in a junction.}
        \label{fig:recmdM2}
    \end{subfigure}
    \begin{subfigure}[t]{0.28\textwidth}
        \includegraphics[width=\textwidth]{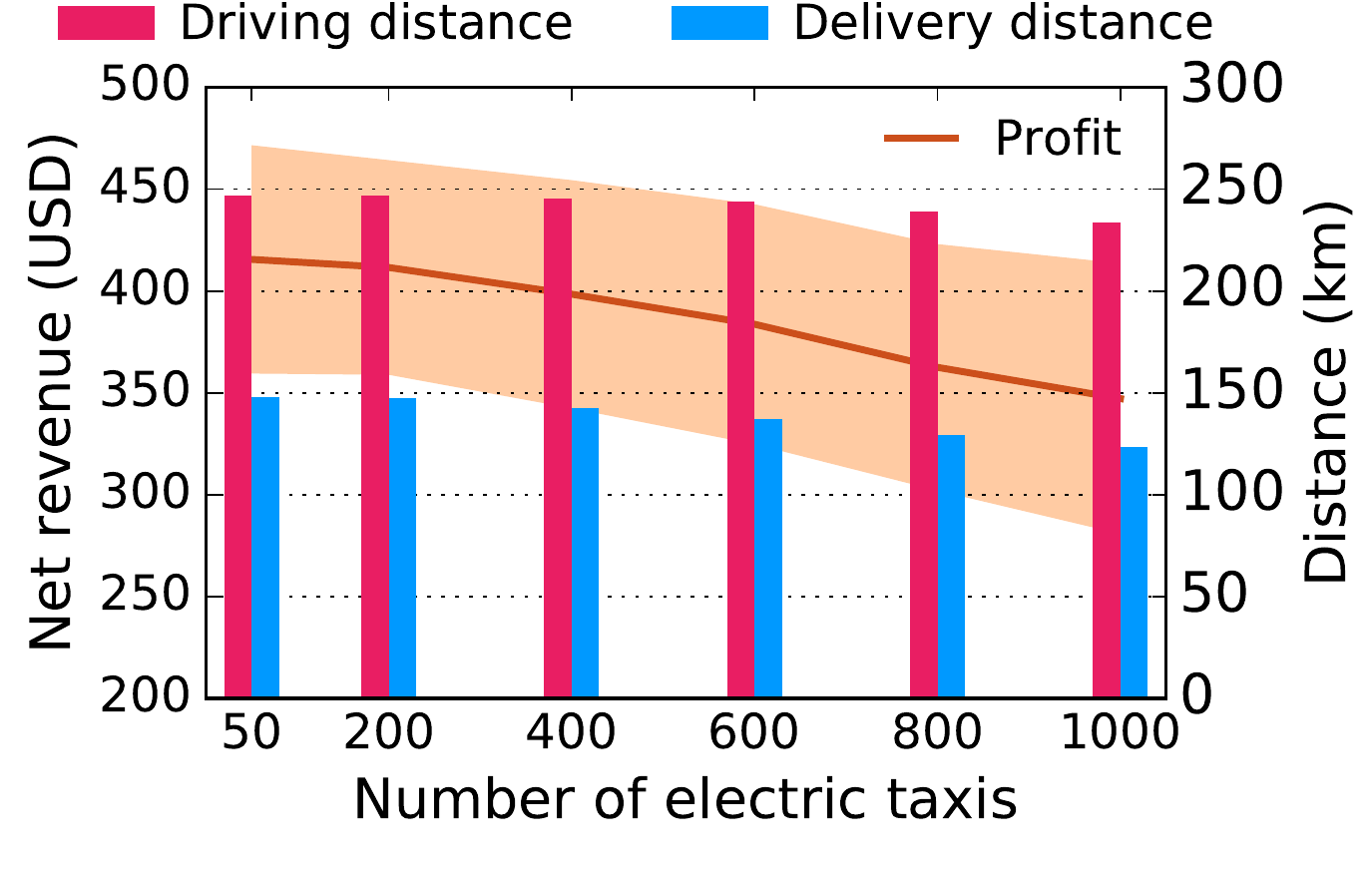} 
        \caption{Estimated net revenues for multiple electric taxis with 70 kWh battery capacity.}
        \label{fig:recmdM1}
    \end{subfigure}    	
    \begin{subfigure}[t]{0.5\textwidth}
        \includegraphics[width=\textwidth]{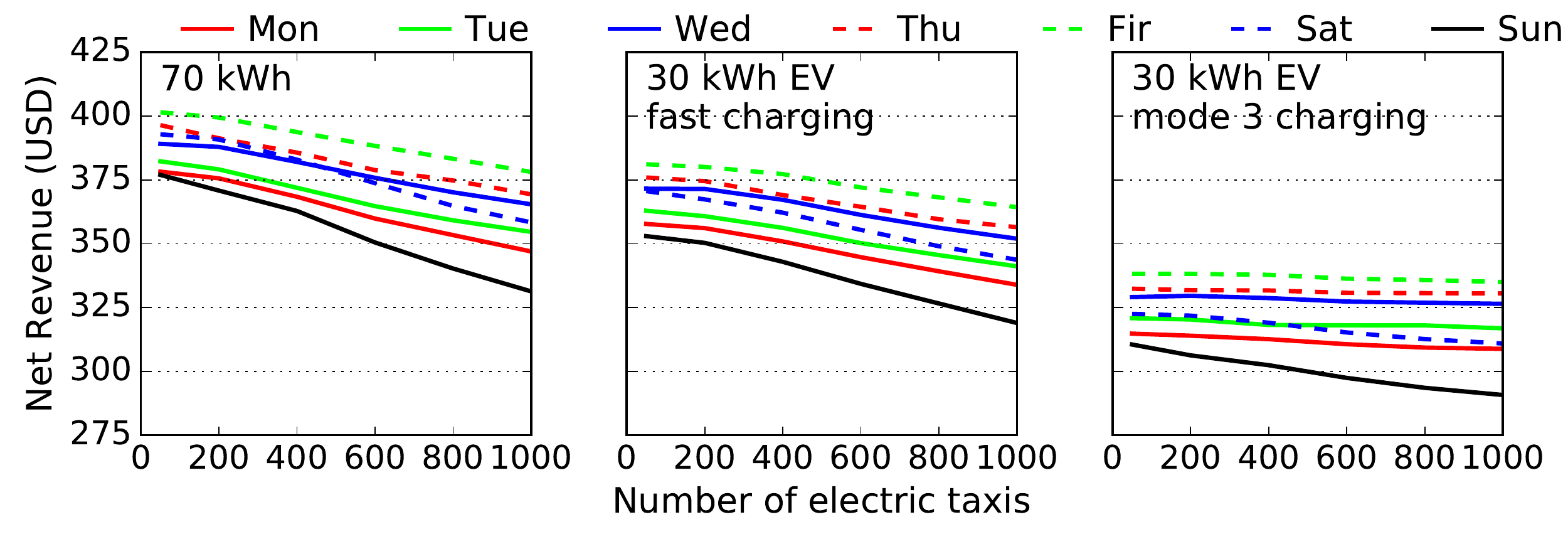} 
        \caption{Estimated net revenues for multiple electric taxis over year 2013.}
        \label{fig:recmdMY}
    \end{subfigure}
    \caption{Estimated net revenues for multiple electric taxis.}
    \label{fig:recmdM}
\end{figure*}

\subsection{Using Historical Data for Prediction}

{\bf Setting:}
The previous basic evaluation of net revenues is based on MDP using the complete knowledge, which requires knowing the pick-up demands and locations in a-priori manner. However, complete information is difficult to obtain in practice. A more practical approach is to use only historical data as training dataset for MDP, and then obtain an optimal policy as a heuristic for other days. In the following, we use the optimal policy of MDP obtained from 6th January to 12th January (i.e., the first week after 1st January) as training data. Then we employ the policy to all morning shifts in the year in the evaluation.

{\bf Observations:}
Fig.~\ref{fig:recmd} shows the estimated net revenue using one-day training data on different days of a week. We observe that the highest net revenue occurs on Friday while the lowest net revenue occurs on Sunday, because of more passengers on Fridays.  The figures also show the benchmark for ICE taxis using historical data (i.e., gray band). In particular, Fig.~\ref{fig:recmd1} shows the net revenue of 70 kWh battery capacity using different  dates in January as training data. We observe that the training data from 6th January performs the best while training data from 8th January performs the worst. A taxi driver can receive 7.5\% higher net revenue using 6th January data.

Fig.~\ref{fig:recmd3} shows the net revenues with the 30 kWh  battery capacity using training data from 6th January. We observe that 12\% higher net revenue can be obtained using fast charging than using mode 3 charging. Electric taxis with 70 kWh  battery capacity can obtain 2.8\% higher than 30 kWh  battery capacity using fast charging.

{\bf Ramifications:}
Using historical data for prediction, instead of complete knowledge, will inevitably reduce the effectiveness. However, this creates a similar effect on ICE taxis that also use historical data. Hence, optimizing taxi service strategies for electric taxis using historical data still achieves comparable net revenues as that of ICE taxis.

\begin{figure*}[!htb]
    \begin{subfigure}[t]{0.23\textwidth}
        \includegraphics[width=\textwidth]{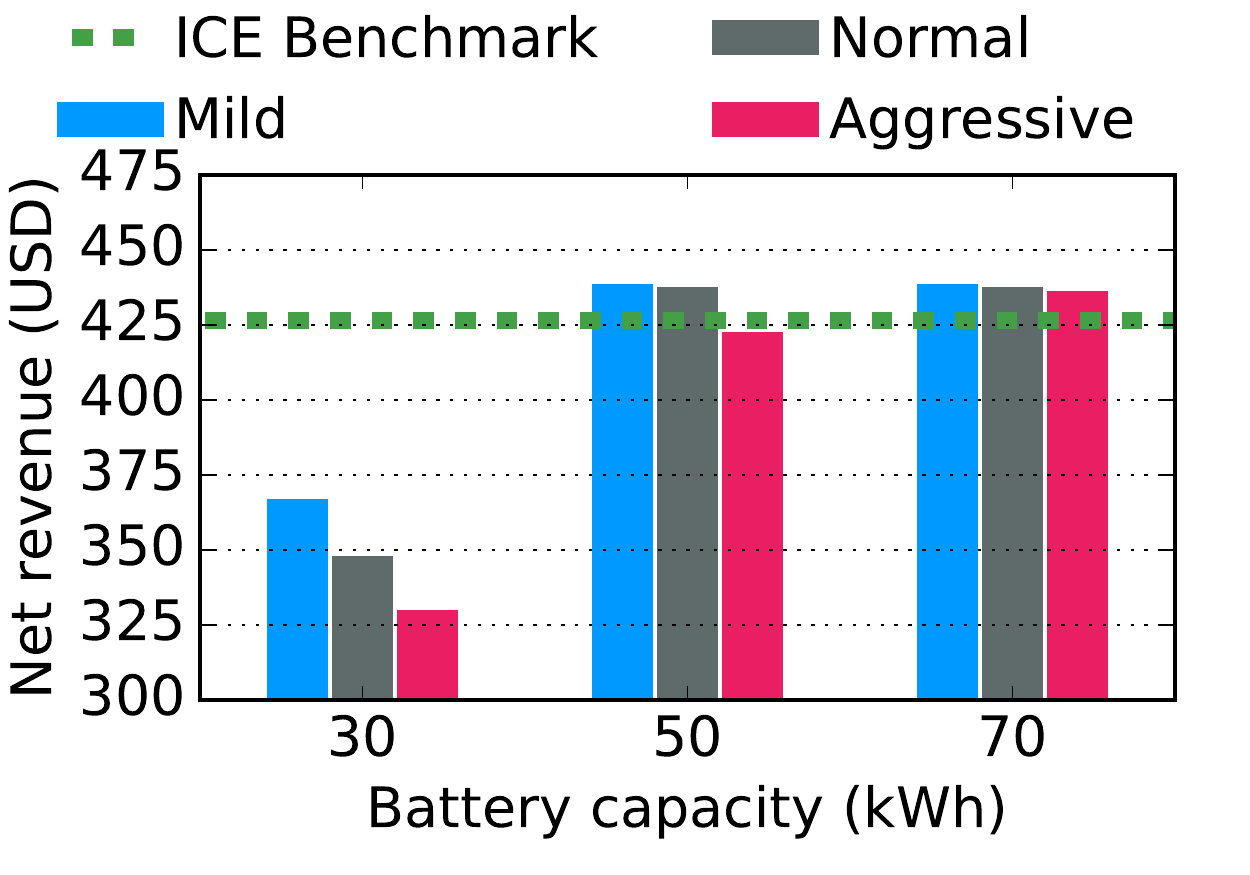} 
        \caption{Estimated net revenues of different driving behaviors using mode 3 charging.}
        \label{fig:EVprofitbehavior}
    \end{subfigure}
 \hspace{-0.em}
    \begin{subfigure}[t]{0.23\textwidth}
        \includegraphics[width=\textwidth]{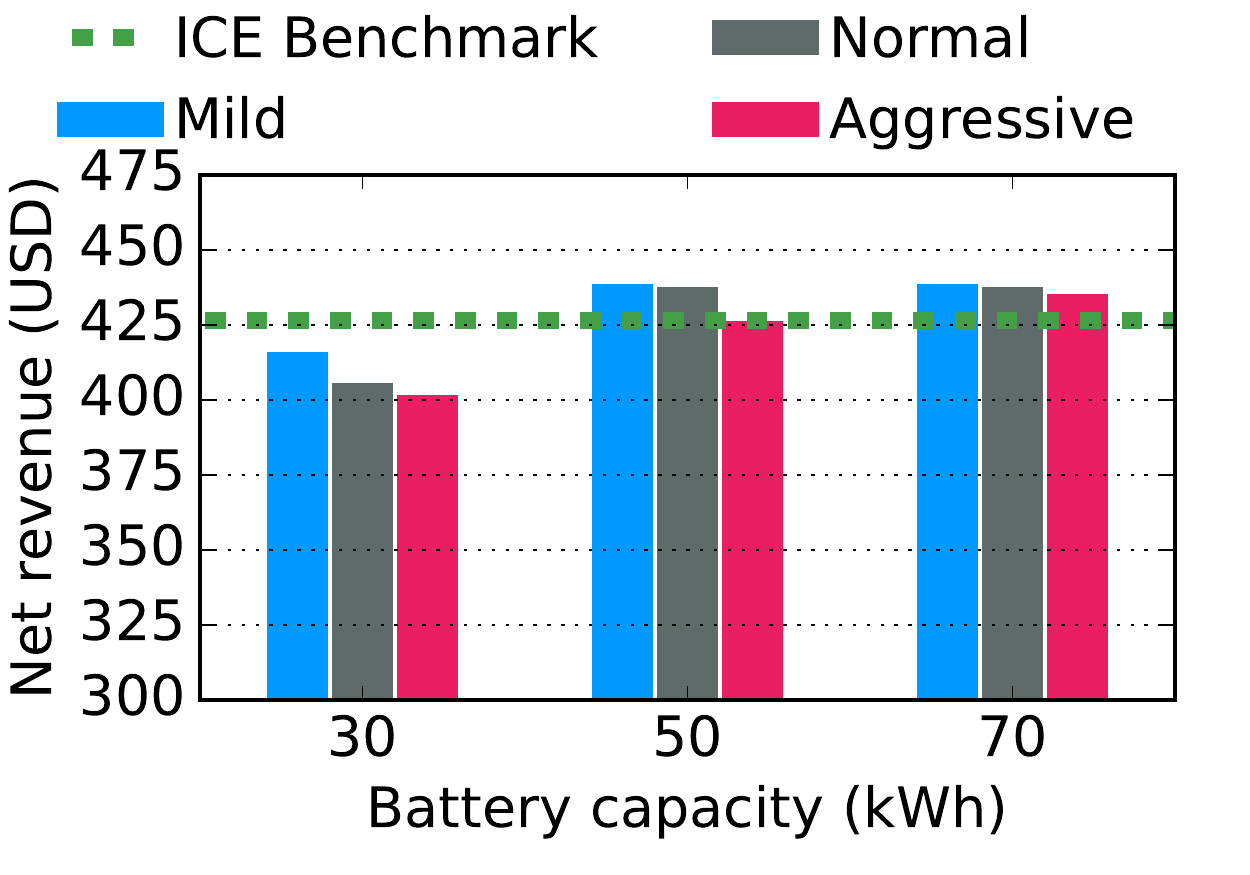} 
        \caption{Estimated net revenues of different driving behaviors using fast charging.}
        \label{fig:EVprofitbehaviorfast}
    \end{subfigure}
 \hspace{-0.em}	
    \begin{subfigure}[t]{0.23\textwidth}
        \includegraphics[width=\textwidth]{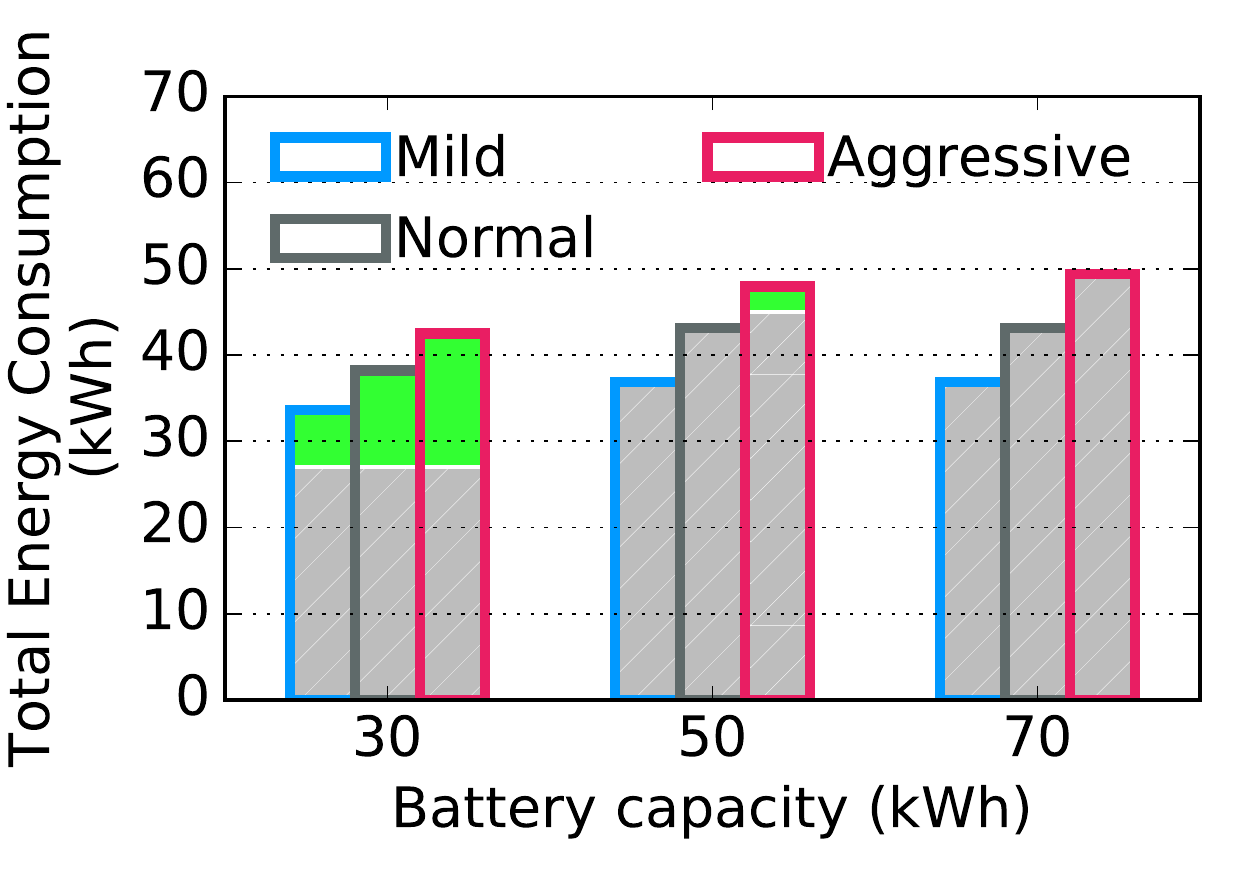} 
        \caption{Energy consumption of different driving behaviors using mode 3 charging.}
        \label{fig:EVprofitbehavioregy}
    \end{subfigure}
 \hspace{-0.em}	
    \begin{subfigure}[t]{0.23\textwidth}
        \includegraphics[width=\textwidth]{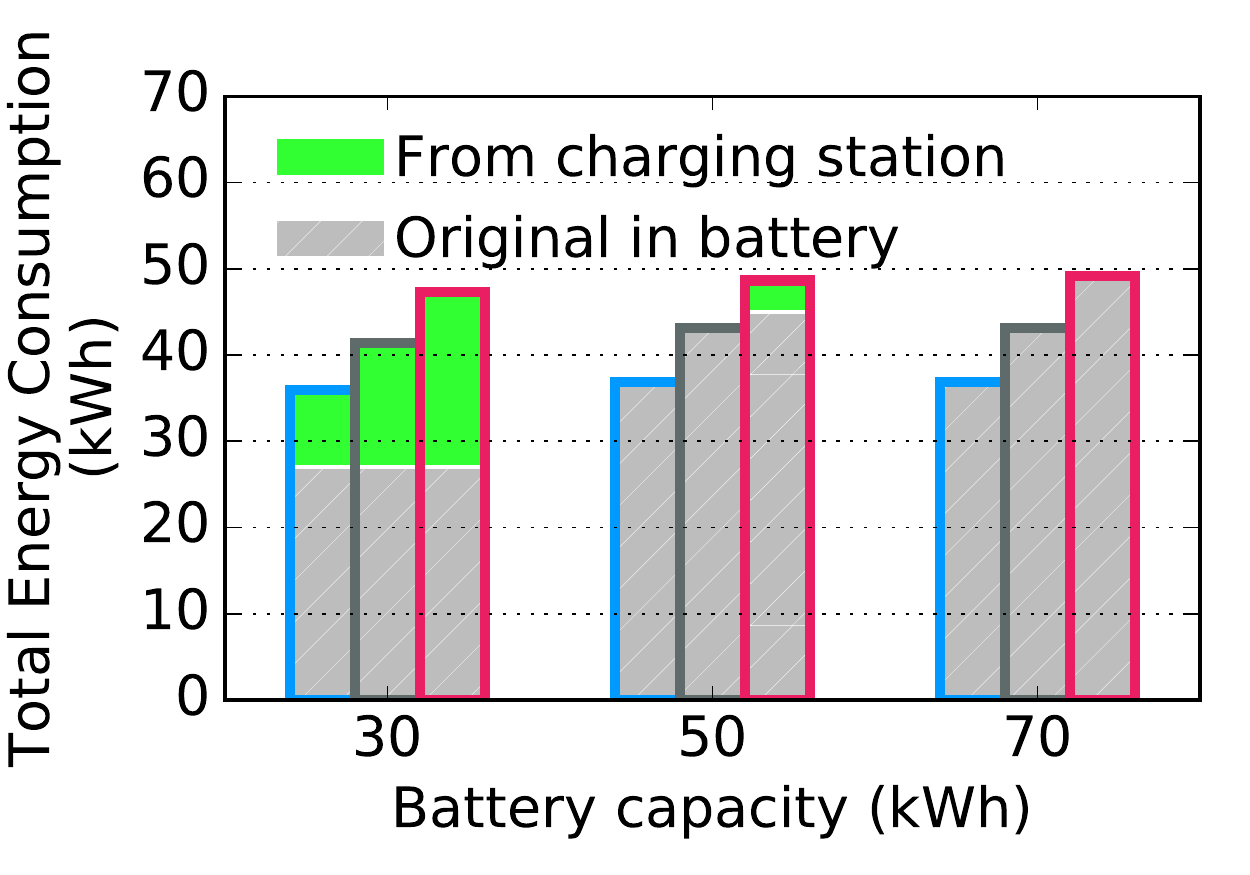} 
        \caption{Energy consumption of different driving behaviors using fast charging.}
        \label{fig:EVprofitbehavioregyfast}
    \end{subfigure}
    \caption{Estimated net revenues and energy consumptions of different driving behaviors.}
    \label{fig:EVprofitbehavioregyAll}
\end{figure*}

\subsection{Multiple Electric Taxis} 
\label{sec:multit}

{\bf Setting:}
The optimal policy from MDP has been previously employed in one single taxi. Next, we employ the optimal policy to multiple taxis. The idea is to allow multiple electric taxis adopt the optimal policy from MDP, while ensuring the number of taxis being sent to each location is constrained. Otherwise, this leads to over-provision of taxis at certain locations. This simple constraint allows us to decouple the individual MDP decisions. Otherwise, considering a large complex problem will be intractable. In practice, we may display the potential net revenue of each junction to the taxi drivers. The junction will become less desirable, when the number of taxis currently present exceeds a certain threshold. Hence, they would not prefer to go to the junction.

We first empirically study the distribution of number of taxis at all the junctions over time from the dataset. We then set of limit of the number of taxis at each junction according to the mean number of taxis at each junction from the dataset. To satisfy the constraint, some electric taxis would need to follow the second-best decisions in the optimal policy. Each taxi state is initialized by the junction and the time according to the dataset. The state of each taxi is tracked and the passenger pick-up probability $P^{\tt{p}}_{t}(i)$ is recomputed using Eqn.~(\ref{eq:pickpr}). We use the optimal policy based on the data of 6th January.

{\bf Observations:}
Fig.~\ref{fig:recmdM2} displays the histogram of number of taxis in a junction. We observe that the number of taxis in each junction is less than 7 by 99\% of time. We set of limit of the number of taxis at each junction according to the mean number of taxis at each junction from the dataset.

Fig.~\ref{fig:recmdM1} shows the net revenues of different numbers of electric taxis using  the optimal policy of MDP on 9 Jan. We observe that the net revenue drops to \$USD 350 when 1000 electric taxis use the optimal policy of MDP. The red bar indicates the total driving distance of the taxis and blue bar indicates the passenger delivery distance. We observe that  the delivery distance drops but the total driving distance remains relatively steady. This implies that the increase of roaming distance is due to a lower passenger pick-up probability. Fig.~\ref{fig:recmdMY} shows the average net revenue of multiple taxis over entire year of 2013. We observe that the highest net revenue occurs on Fridays while the lowest occurs on Sundays. We also observe that the net revenue is less affected by the number of taxis when mode 3 charging is used. This is because that the electric taxis require frequent recharging, which may result in less available taxis, and hence, a higher pick-up probability.

{\bf Ramifications:}
If the optimal policy of MDP is deployed up to 1000 electric taxis, then the net revenues will decrease, as a result of diminishing advantage of computerized service strategies. These 1000 taxi drivers can still earn as top 1.7\% among traditional taxi drivers without computerized service strategies.

\subsection{Considering Driving Behavior}

{\bf Setting:}
Driving behavior plays an important role in energy consumption of vehicles. Aggressive driving behavior results in more energy consumption. Furthermore, higher energy consumption rate induces more frequent recharging of EVs, which reduces the net revenues of the taxi drivers. We study three classes of driving behaviors: i) mild drivers ($\beta=0.8$), ii) normal drivers ($\beta=1$), and iii) aggressive drivers ($\beta=1.2$).

{\bf Observations:}
Fig.~\ref{fig:EVprofitbehavioregyAll} shows the estimated net revenues of different driving behaviors for morning shifts. Fig.~\ref{fig:EVprofitbehavior} shows the estimated net revenues of different driving behaviors using mode 3 charging. Mild drivers can receive 14\% higher net revenue than aggressive drivers when driving 30 kWh Leaf using mode 3 charging. However, the net revenue is less affected by different drivers when the battery capacity is sufficiently large to eliminate recharging during a shift. Fig.~\ref{fig:EVprofitbehaviorfast} shows the estimated net revenues using fast charging. We observe that the net revenue is also less affected by different driving behaviors because of shorter recharging duration. Fig.~\ref{fig:EVprofitbehavioregyAll} shows the energy consumption of different driving behaviors. We observe that aggressive drivers consume around 11 kWh more energy than mild drivers.

{\bf Ramifications:}
Although the aggressive drivers consumes 20\% more energy which only results in \$USD2.2 difference for morning shifts. The result shows that the driving behavior only has a higher impact on the net revenue when the battery capacity is insufficient to eliminate recharging during a shift.

\subsection{Considering Different Gas Prices}

{\bf Setting:}
To complete the study of viability of electric taxis, we provide a study of ICE taxis' net revenue under different gas prices. Note that the current gas price in USA is around USD\$2.5 per gallon, while the current gas price in China is around 7.2 RMB per liter, which is equivalent to \$4.5 USD per gallon. We analyze the outcomes of three different gas prices (i.e., \$2.5 USD/G, \$3.5 USD/G, \$4.5 USD/G) considering the optimal policy of MDP for an ICE taxi.

\begin{figure}[H]
\center
    \includegraphics[width=0.44\textwidth]{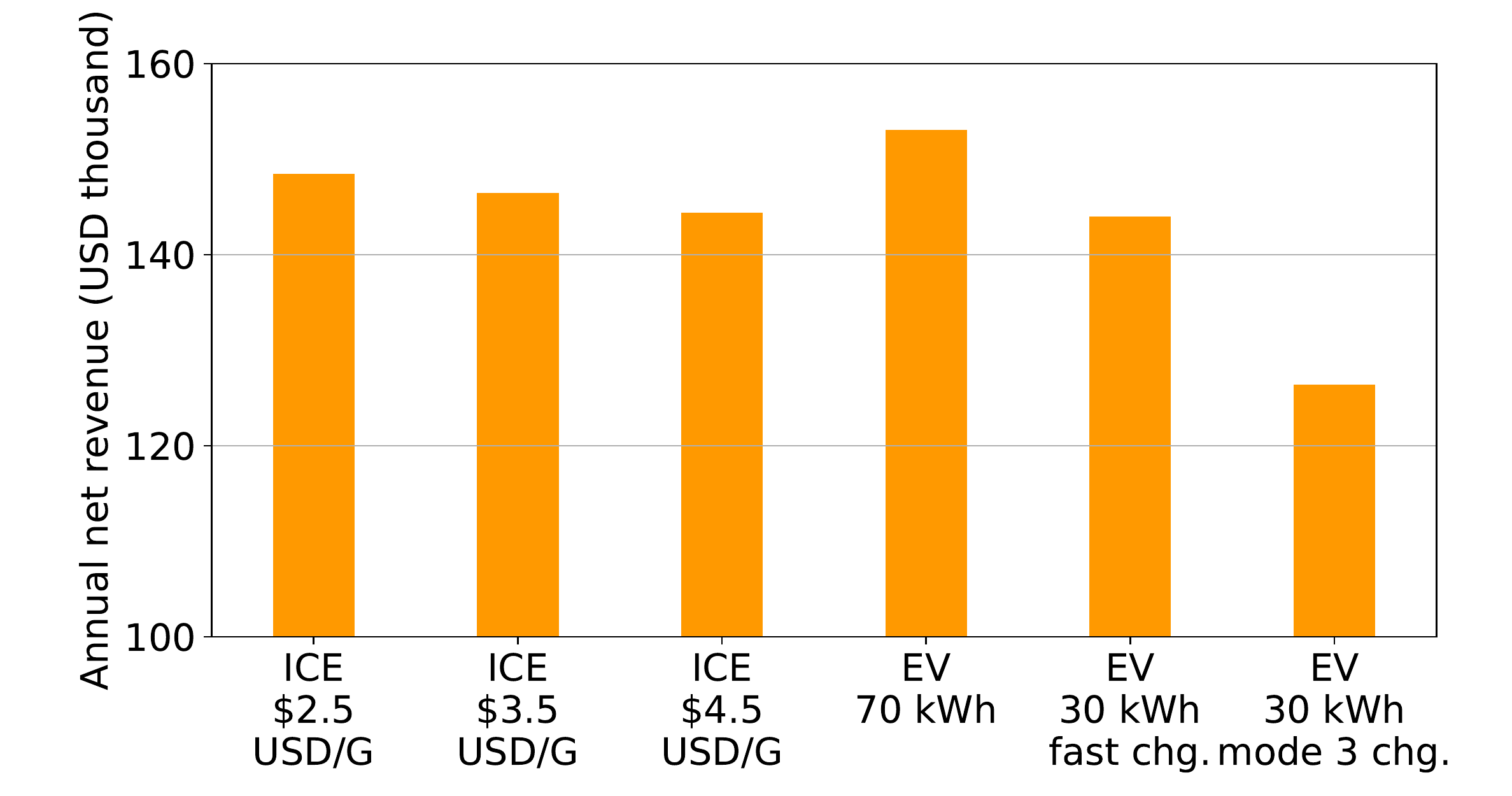}
    \caption{Annual net revenues under different gas prices.}
    \label{fig:gaspriceanalysis}
\end{figure}

{\bf Observations:}
Fig.~\ref{fig:gaspriceanalysis} compares the annual net revenues of ICE taxi under different gas prices, with that of electric taxis using different charging options. We observe that the annual net revenue with gas price \$2.5 USD/G (i.e., the leftmost bar) is slightly higher (about USD\$ 4000 higher) than that with gas price \$4.5 USD/G. We also observe that the comparable net revenue can be achieved by 30 kWh EV with fast charging when gas price increases to \$4.5 USD/G. However, the annual net revenue of 30 kWh EV with mode 3 charging is much lower (about 14\% lower), even when the gas price is high.

{\bf Ramifications:}
We observe that when the gas price increases, ICE taxi becomes a less attractive option since its net revenue decreases. The net revenue of 70 kWh EV is around 3\% higher than ICE taxi when gas price is \$2.5 USD/G, while it is around 6.6\% higher than ICE taxi when gas price increases to \$4.5 USD/G.

\begin{figure*}[!htb]
    \begin{subfigure}[t]{0.24\textwidth}
        \includegraphics[width=\textwidth]{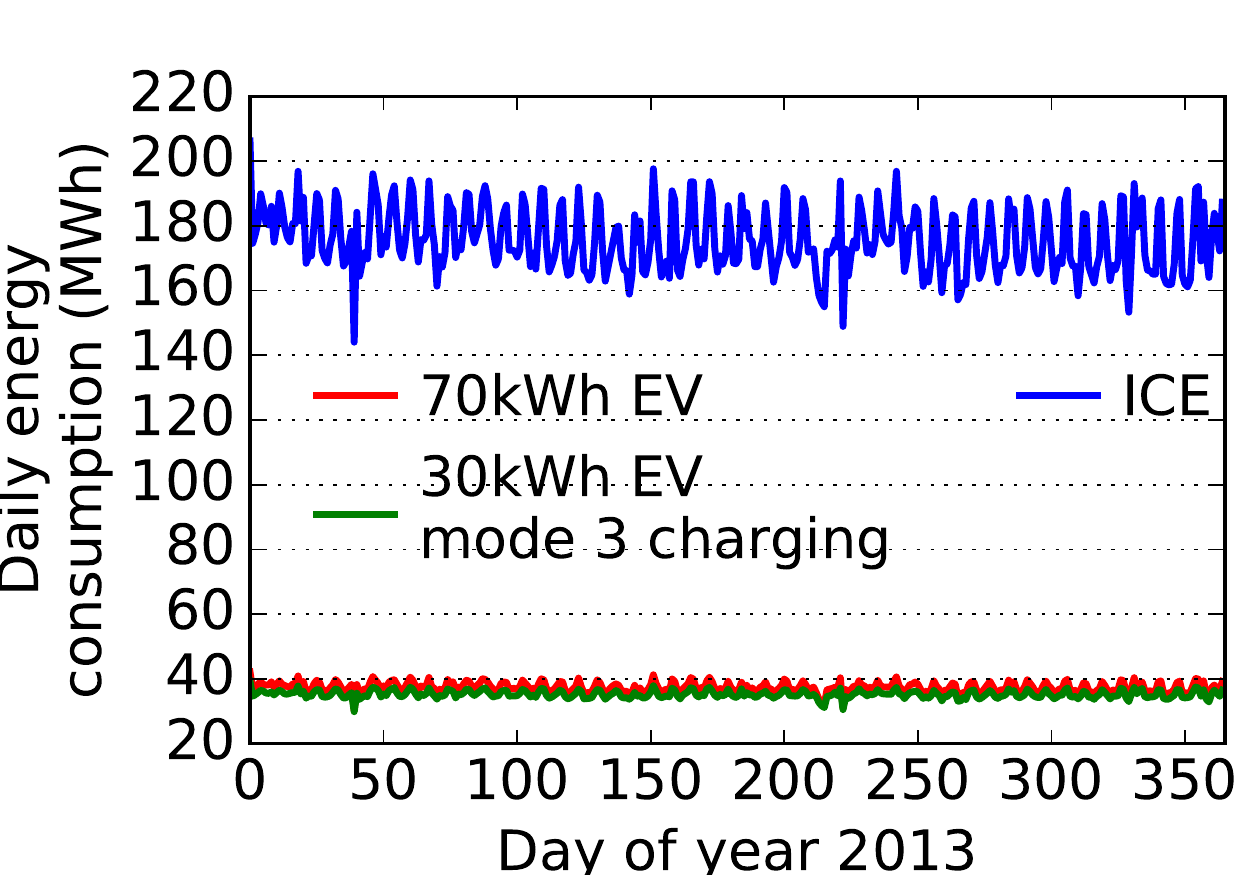} 
        \caption{Daily energy consumption of morning shifts.}
        \label{fig:dailyegy}
    \end{subfigure} ~
    \begin{subfigure}[t]{0.21\textwidth}
        \includegraphics[width=\textwidth]{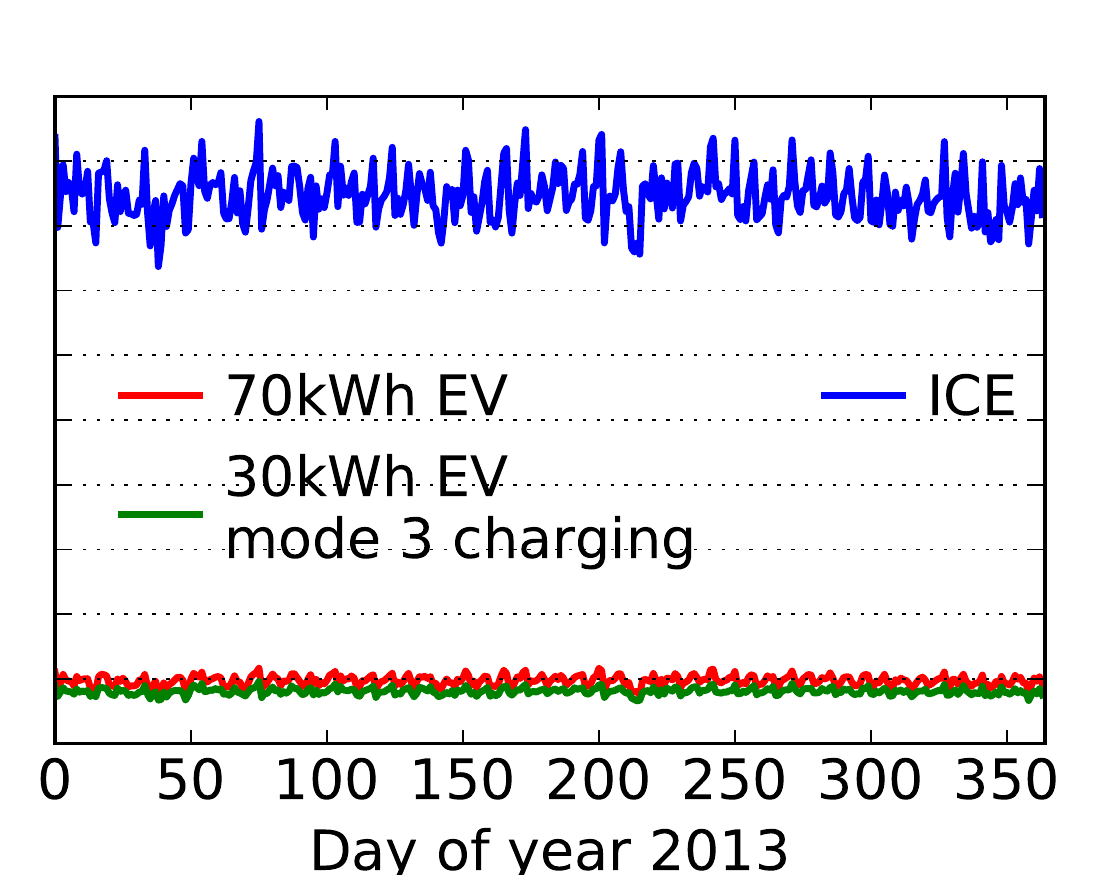} 
        \caption{Daily energy consumption of night shifts.}
        \label{fig:dailyegyN}
    \end{subfigure}   ~
    \begin{subfigure}[t]{0.24\textwidth}
        \includegraphics[width=\textwidth]{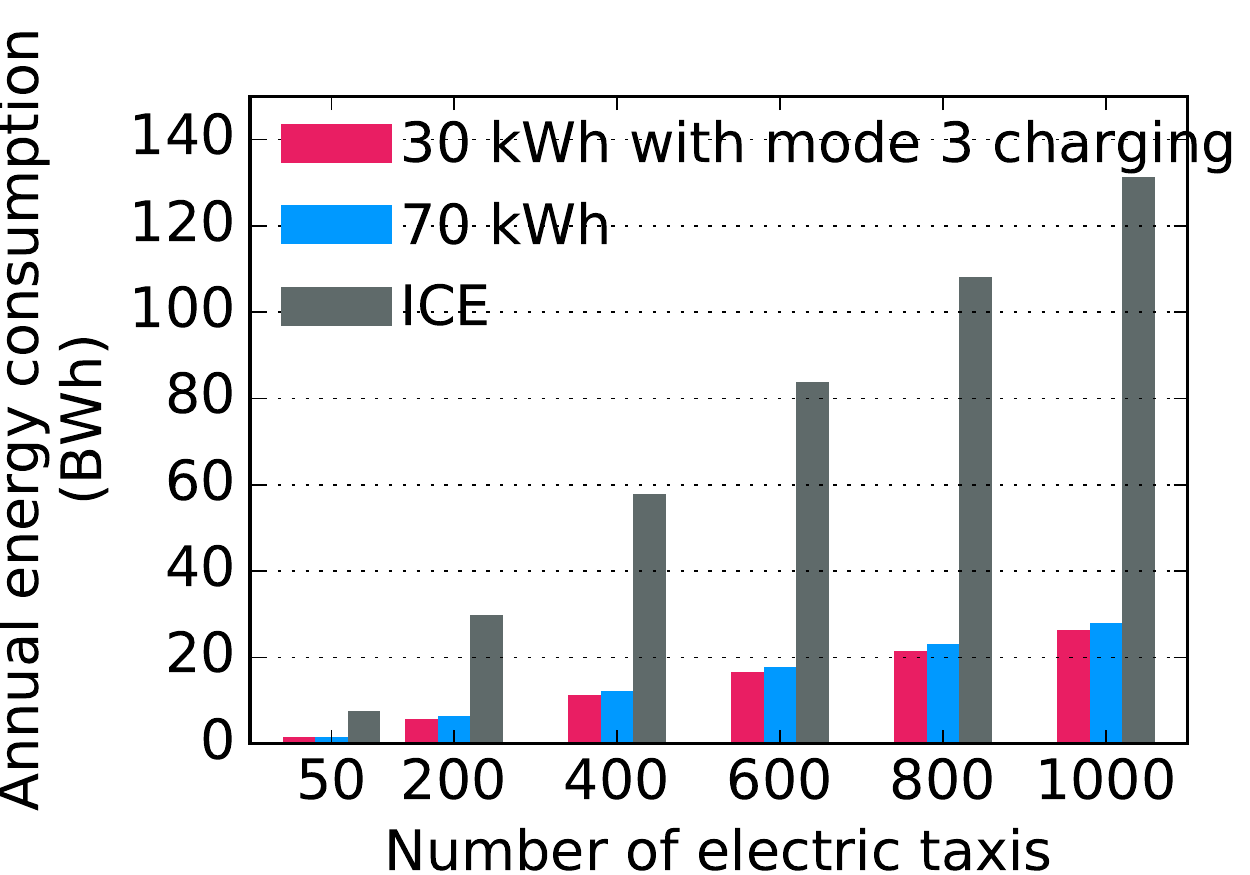} 
        \caption{Annual energy consumption of different numbers of electric taxis.}
        \label{fig:annualegynum}
    \end{subfigure} ~
    \begin{subfigure}[t]{0.24\textwidth}
        \includegraphics[width=\textwidth]{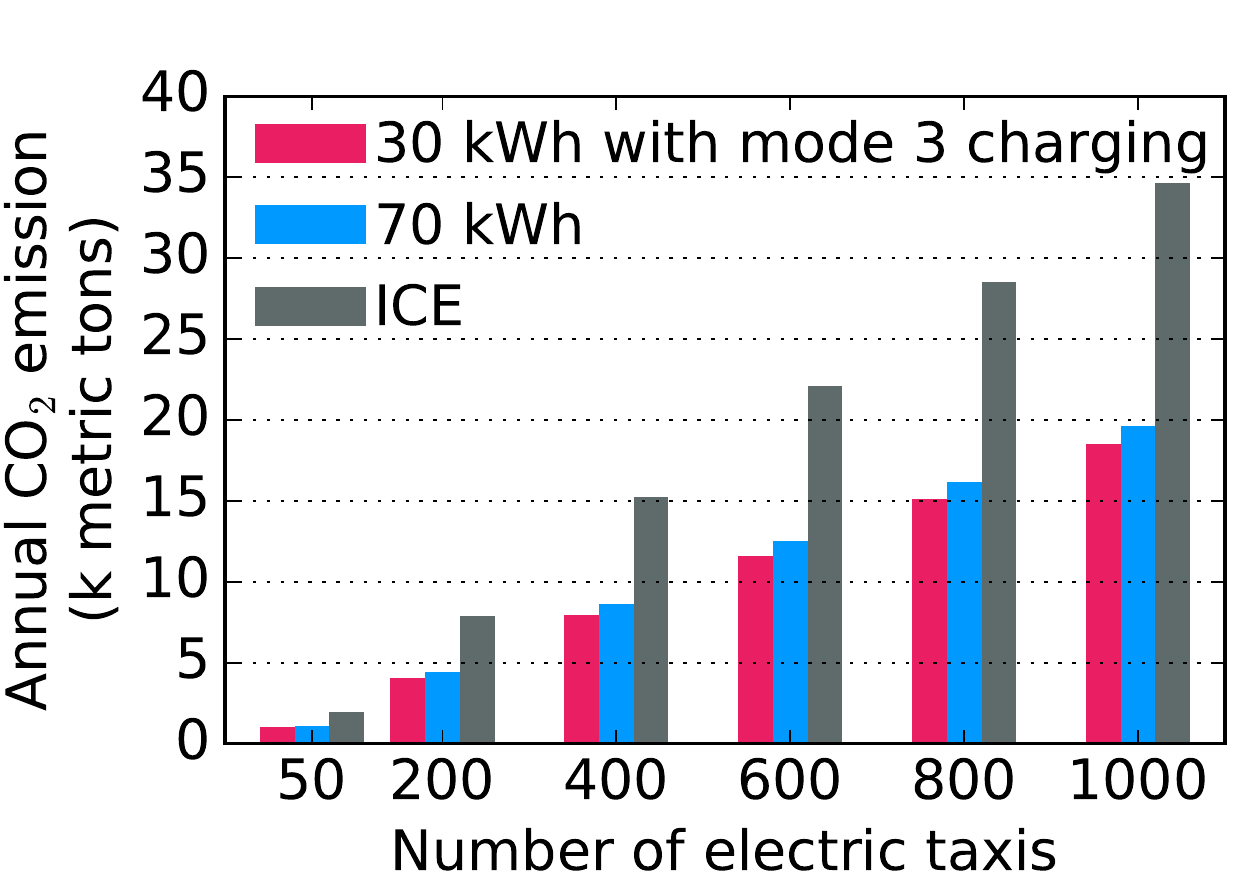} 
        \caption{Annual CO$_2$ emission of different numbers of electric taxis.}
        \label{fig:annualco2num}
    \end{subfigure}
    \caption{Daily and annual energy consumption and CO$_2$ emission.}
    \label{fig:annualegy}
\end{figure*}

\subsection{Annual Evaluation of NYC Taxi Trip Dataset}
\label{sec:emi}

\subsubsection{Net Revenue Evaluation}
\

\begin{figure}[H]
    \includegraphics[width=.5\textwidth]{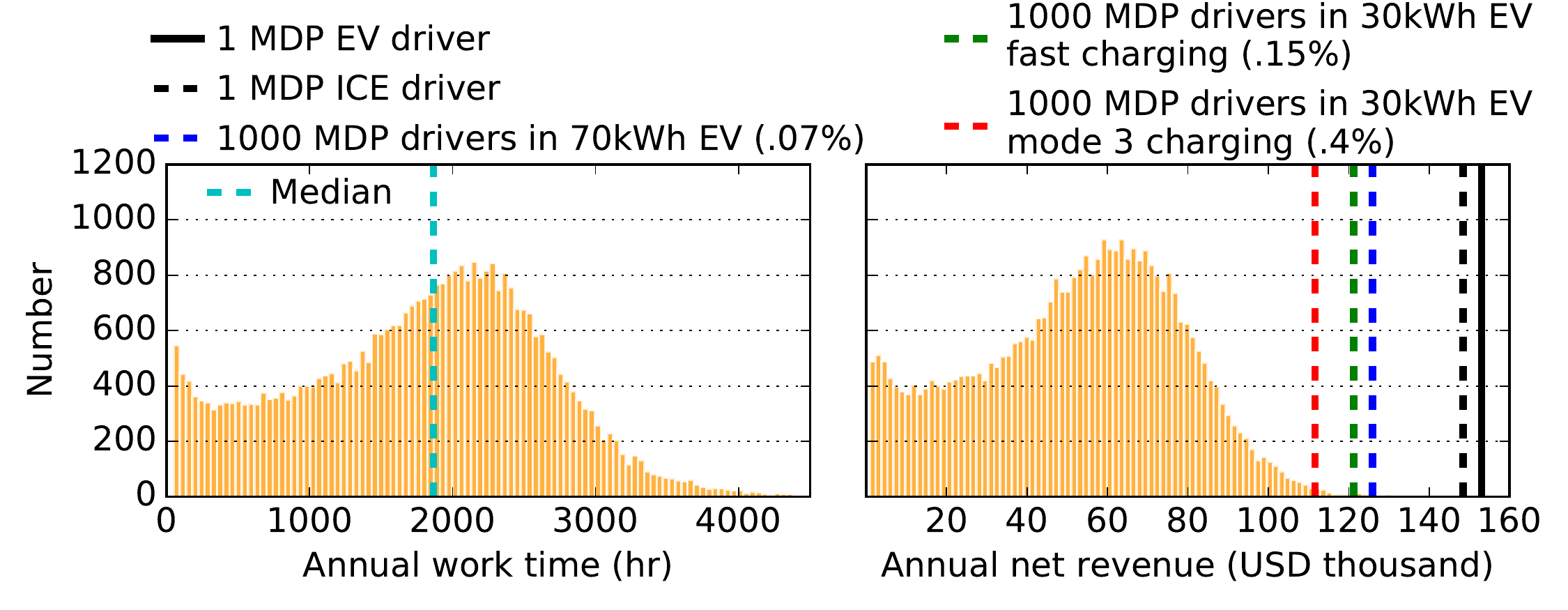}  
    \caption{Annual working hours and estimated net revenues of taxi drivers in 2013.}
    \label{fig:annualprofitanalysis}
\end{figure}

{\bf Setting:}
We employ the optimal policy from 6th January to different numbers of electric taxis and estimate the annual net revenues. Fig.~\ref{fig:annualprofitanalysis} shows the distribution of annual working hours, we observe that the median annual working hour is around 1800 hours, but many drivers work more than 4300 hours, equivalent to working almost 12 hours a day. Therefore, we consider taxi drivers working every morning shift (i.e., 4380 working hours) to estimate their net revenues.

{\bf Observations:}
The right figure in Fig.~\ref{fig:annualprofitanalysis} shows the estimated net revenues of different taxi drivers using the optimal policy of MDP. There are some observations:
\begin{itemize}
\item The case of one electric taxi driver using the optimal policy of MDP can earn 3\% higher than that of one ICE taxi driver.
\item The average net revenue of case of 1000 electric taxis with 70 kWh battery is ranked top 0.07\% among traditional taxi drivers without computerized service strategy.
\item The average net revenue of case of 1000 electric taxis with 30 kWh battery using mode 3 charging is ranked top 0.4\% among traditional taxi drivers without computerized service strategy.
\end{itemize}
The results shows that the optimal policy of MDP can enable electric taxi drivers to make comparable revenues as traditional taxi drivers.

\subsubsection{Carbon Emission Evaluation}
\ 

{\bf Setting:}
Besides of net revenues as economic motivation, an important benefit is the reduction of carbon emission by switching from ICE taxis to electric taxis. Although electric taxis do not produce tailpipe emissions, the electricity grid to recharge the battery may still produce emissions. In this section, we estimate the CO$_2$ emission of electric taxis, as compared with ICE taxis, with  computerized service strategy optimization.
The CO$_2$ emission factors of electricity and gasoline are obtained from eGrid of Long Island \cite{epa}:
\begin{itemize}
\item Emission factor of electricity:
0.7007 kg/kWh
\item Emission factor of gasoline:
2.348 kg/liter
\end{itemize}

{\bf Observations:}
We consider taxis working in all shifts. Fig.~\ref{fig:dailyegy} shows the daily energy consumption of 1000 taxis for morning shifts, while Fig.~\ref{fig:dailyegyN} shows the daily energy consumption for night shifts.  We use miles per gallon gasoline equivalent to convert the consumed gasoline to kWh (i.e., 1 gallon of gasoline equals to 33.7 kWh). Fig.~\ref{fig:annualegynum} shows the annual energy consumption of different numbers of electric taxis. We observe that ICE taxis consume around 4 times more energy than electric taxis. Fig.~\ref{fig:annualco2num} shows the corresponding CO$_2$ emissions of different numbers of electric taxis. We observe that  up to 15 thousand metric tons CO$_2$ (equal to 1560 home's energy use for one year) can be saved by replacing 1000 ICE taxis by electric taxis.

\section{Conclusion}

In this paper, we employ Markov Decision Process to model computerized taxi service strategy and optimize the strategy for taxi drivers considering electric taxi operational constraints. We evaluate the effectiveness of the optimal policy of Markov Decision Process using a big data study of real-world taxi trips in New York City. The optimal policy can be implemented in an intelligent recommender system for taxi drivers. This becomes more viable especially due to the advent of autonomous vehicles. Our evaluation shows that computerized service strategy optimization allows electric taxi drivers to earn comparable net revenues as ICE drivers, who also employ computerized service strategy optimization, with at least 50 kWh battery capacity. Hence, this sheds light on the viability of electric taxis.

\bibliographystyle{IEEEtran}
\bibliography{paperbib}

\section*{Acknowledgment} 
The authors would like to thank Srinivasan Keshav and Sgouris Sgouridis for helpful suggestions and discussions.

\end{document}